\documentclass[12pt]{scrartcl}
\usepackage{amsmath,amssymb}
\usepackage{graphicx}
\usepackage{cite}
\usepackage{here}
\usepackage{subcaption}
\usepackage[UKenglish]{babel}

\allowdisplaybreaks
\begin{document}

\thispagestyle{empty}
\noindent \today
\vspace{3.0cm}
\begin{center}
\LARGE{\textbf{Forward-backward asymmetry \\ in the gauge-Higgs unification \\ at the International Linear Collider}}
\end{center}

\vspace{1.0cm}
\begin{center}
{Shuichiro Funatsu}
\vspace{0.5cm}

\textit{\small 
	Institute of Particle Physics and Key Laboratory of Quark and Lepton Physics (MOE), \\
	Central China Normal University, Wuhan, Hubei 430079, China}
\end{center}

\vspace{2.0cm}
\begin{abstract}
	The signals of the $SO(5)\times U(1)$ gauge-Higgs unification model at the International Linear Collider are studied. 
	In this model, Kaluza-Klein modes of the neutral gauge bosons affect fermion pair productions. 
	The deviations of the forward-backward asymmetries of the $e^+e^-\to \bar{b}b$, $\bar{t}t$ processes from the standard model predictions 
	are clearly seen by using polarised beams. 
	The deviations of these values are predicted for two cases, the bulk mass parameters of quarks are positive and negative case. 
\end{abstract}

\newpage
\section{Introduction}
After the discovery of the Higgs boson, 
search for new physics in the electroweak sector is one of the most important topics of the particle physics. 
For this purpose, high-precision measurements of the electroweak sector are necessary. 
The International Linear Collider (ILC) has the capabilities for the high-precision measurements 
and enables us to test the standard model (SM)~\cite{Baer:2013cma,Asner:2013psa,Aihara:2019gcq,Bambade:2019fyw}. 
New physics at the ILC are predicted by many alternative models 
such as the Higgs portal dark matter models~\cite{Kanemura:2011nm,Ko:2016xwd,Kamon:2017yfx}, 
two-Higgs doublet models~\cite{Barger:2013ofa,Kanemura:2014bqa,Grzadkowski:2016lpv}, 
Georgi-Machacek model~\cite{Chiang:2015rva}, 
supersymmetric models~\cite{Berggren:2013vfa,Cahill-Rowley:2014wba,Heinemeyer:2015qbu,Heinemeyer:2016wey},
littlest Higgs model~\cite{Cho:2004xt,Yue:2005nc}, 
universal extra dimensional model~\cite{Riemann:2005es},
warped extra dimensional models~\cite{Hewett:2002fe,Djouadi:2006rk,Angelescu:2017jyj},
composite Higgs models~\cite{Contino:2013gna,Barducci:2013ioa,Kanemura:2016tan}
and other models~\cite{Osland:2009dp}.

The gauge-Higgs unification (GHU) models~\cite{Hosotani,Davies,Hatanaka,Csaki:2002ur,Burdman:2002se,Medina:2007hz,Funatsu:2013ni,Yamamoto:2013oja,Funatsu:2014fda,Funatsu:2014tka,Hosotani:2015hoa,Funatsu:2015xba,Yamatsu:2015oit,Kojima:2017qbt,Sakamura:2006rf,Hosotani:2008tx,Hosotani:2008by, Matsumoto:2016okl,Furui:2016owe,Funatsu:2016uvi,Funatsu:2017nfm,Hosotani:2017hmu,Hosotani:2017edv,Hasegawa:2018jze,Yoon:2017cty,Yoon:2018vsc,Yoon:2018xud,Funatsu:2019xwr} are
alternative models of the SM, 
in which the Higgs boson appears as an extra-dimensional component of the higher-dimensional gauge boson. 
Hence the Higgs sector is governed by the gauge principle 
and the Higgs boson mass is protected against the radiative corrections in the GHU models. 
The Higgs boson is massless at the tree-level and acquires the finite mass by the radiative corrections~\cite{Hosotani,Davies,Hatanaka}. 
Note that also the Yukawa interactions in the GHU models are the gauge interactions of the extra-dimensional component of the gauge boson and the fermions.  
The phenomenologically most well-studied GHU model is the $SO(5)\times U(1)$ GHU models~\cite{
	Medina:2007hz,
	Sakamura:2006rf,Hosotani:2008tx,Hosotani:2008by,
	Funatsu:2013ni,Funatsu:2014fda,Funatsu:2014tka,Funatsu:2015xba,Funatsu:2016uvi,Funatsu:2017nfm,Yoon:2017cty,Yoon:2018vsc,Yoon:2018xud,Funatsu:2019xwr}, 
which are defined on the warped metric~\cite{Randall:1999ee}. 
In the $SO(5)\times U(1)$ GHU models, the Higgs boson mass is protected by the custodial symmetry and the Higgs boson potential is calculable. 
One of the features of the warped extra dimensional models is that the first Kaluza-Klein (KK) excited states of the gauge bosons have the large asymmetry between the couplings with the left-handed and the right-handed fermions. 
The $SO(5)\times U(1)$ GHU models have the same feature. 

There are several kinds of the $SO(5)\times U(1)$ GHU models 
depending on symmetry breaking patterns and the embedding patterns of the quarks and leptons. 
In the $SO(5)\times U(1)$ GHU model discussed here, 
the quarks and leptons are embedded in the \textbf{5} representation of $SO(5)$. 
Then the Yukawa, $HWW$ and $HZZ$ couplings are suppressed by $\cos\theta_H$ from the SM values, 
where $\theta_H$ is the Wilson line phase. 
The KK excited states of the neutral gauge bosons appear as the so-called $Z'$ bosons. 
Besides the KK excited states of the photon $\gamma^{(n)}$ and that of the $Z$ boson $Z^{(n)}$, 
that of the $SU(2)_R$ gauge boson $Z_R^{(n)}$, which does not have a zero mode, 
exist as the neutral gauge bosons. 
Therefore the $\gamma^{(n)}$, $Z^{(n)}$ and $Z_R^{(n)}$ appear as the $Z'$ bosons. 
From the result at the Large Hadron Collider, the parameter region is constrained to be $\theta_H\lesssim0.1$, 
where $m_\text{KK}\gtrsim 8$~TeV~\cite{Funatsu:2016uvi}. 
For this region, the contributions of the KK excited states to the decay widths, 
$\Gamma(H\to\gamma\gamma)$ and $\Gamma(H\to Z\gamma)$ are less than 0.2\%, thus negligible~\cite{Funatsu:2013ni,Funatsu:2015xba}. 
The decay widths $\Gamma(H\to WW)$, $\Gamma(H\to ZZ)$, $\Gamma(H\to \bar{q}q)$, $\Gamma(H\to l^+l^-)$, $\Gamma(H\to\gamma\gamma)$ and $\Gamma(H\to Z\gamma)$ are approximately suppressed by the common factor $\cos^2\theta_H$ at the leading order. 
Hence the branching ratio of the Higgs boson in this model is almost equivalent to that in the SM. 
At the ILC 500 GeV, the branching ratios of the processes $H\to WW$, $H\to ZZ$, $H\to \bar{b}b$, $H\to \bar{c}c$ and $H\to\tau^+\tau^-$ 
are measured at the $O(1)\%$ accuracy~\cite{Asner:2013psa}. 
The branching ratios of the Higgs boson are consistent between the GHU model and the SM at the ILC. 

At the ILC, it is possible to measure effects of new physics 
on the the cross sections and the forward-backward asymmetries of the $e^+ e^- \to f\bar{f}$ processes~\cite{Cho:2004xt,Yue:2005nc,Hewett:2002fe,Djouadi:2006rk,Amjad:2015mma,Bilokin:2017lco,Bernreuther:2017cyi,Richard:2018zhl}. 
The forward-backward asymmetries are important observables for indirect search of physics beyond the SM 
as shown in the studies of that of top quark production at the Tevatron and Large Hadron Collider~\cite{Jung,Degrande:2010kt}. 
In this paper, the effects of $Z'$ bosons in the GHU model on these values are shown. 
The same topic is studied previously~\cite{Funatsu:2017nfm} 
and clear signals are predicted although $Z'$ masses are much larger than the centre-of-mass of the ILC. 
However in the previous study, the bulk mass parameters are assumed to be positive although negative values are also allowed. 
Thus the negative region of the bulk mass parameters is also checked in this study. 
The following calculations are done at the tree-level without the one-loop effective potential of the Higgs boson, 
so the corrections from the strong interaction are not included. 

This paper is constructed as follows. 
In Section~2, the model is shortly introduced. 
In Section~3, the parameters and the couplings and decay widths of the $Z'$ bosons are shown. The bulk mass parameter dependence of the fermion mode functions are also shortly reviewed. 
In Section~4, the cross sections and the forward-backward asymmetries at the ILC are shown. 
In section~5, the results are summarised.

\section{Model}
There are several types of the $SO(5)\times U(1)$ GHU models 
depending on the symmetry breaking patterns and the embedding patterns of the quarks and leptons. 
The $SO(5)\times U(1)$ GHU model discussed in this paper is constructed as follows. 
The model is defined on the warped spacetime~\cite{Randall:1999ee}.
The metric is 
\begin{align}
	ds^2= g_{MN} dx^M dx^N =
	e^{-2\sigma(y)} \eta_{\mu\nu}dx^\mu dx^\nu+dy^2,
	\label{5D-metric}
\end{align} 
($0 \le |y| \le +L$) where $k$ is the AdS curvature. 
The action has the $SO(5)\times U(1)_X$ local symmetry. 
The bulk action is given by
\begin{align}
	S_\text{bulk} &= \int d^5 x \sqrt{-G} \biggl\{
	-\frac{1}{4}\text{tr} F_{MN}^{(A)} F^{(A) MN}
	-\frac{1}{4} F^{(B)}_{MN} F^{(B) MN}
	-\frac{1}{4}\text{tr} F_{MN}^{(G)} F^{(G) MN}
	\nonumber\\& 
	+ \frac{1}{2\xi_{(A)}} \left(f_\text{gf}^{(A)}\right)^2
	+ \frac{1}{2\xi_{(B)}} \left(f_\text{gf}^{(B)}\right)^2
	+ \frac{1}{2\xi_{(G)}} \left(f_\text{gf}^{(G)}\right)^2
	+ \mathcal{L}_{GH}^{(A)}
	+ \mathcal{L}_{GH}^{(B)}
	+ \mathcal{L}_{GH}^{(G)}
	\nonumber\\
	\label{bulk-action}
	&
	+ \sum_{g=1}^3 \sum_{a=1}^4 
	\bar{\Psi}^{g}_{a} \mathcal{D}(c^{g}_{a}) \Psi^{g}_{a}
	+ \sum_{i=1}^{N_F} \bar{\Psi}_{F_i} \mathcal{D}(c_{F_i}) \Psi_{F_i} \biggr\}, \\
	\mathcal{D}(c_a) &\equiv \Gamma^A e_{A}{}^{M}  
	\bigg(\partial_M + \frac{1}{8} \Omega_{MBC} [\Gamma^B, \Gamma^C]- c_a k \epsilon(y) \nonumber\\
	&\hspace{6em}- i g_A A_M - i g_B Q_{X,a}B_M - i g_C Q_{C,a}G_M \bigg),
\end{align}
where 
$A_M, B_M$ and $G_M$ are the $SO(5)$, $U(1)_X$ and $SU(3)_C$ gauge fields,  
$F_{MN}^{(A)} = \partial_M A_N - \partial_N A_M - i g_A [A_M, A_N]$, 
$F_{MN}^{(B)} = \partial_M B_N - \partial_N B_M$ and 
$F_{MN}^{(G)} = \partial_M G_N - \partial_N G_M - i g_C [G_M, G_N]$, 
$g_A$, $g_B$ and $g_C$ are the five-dimensional gauge couplings of $SO(5)$, $U(1)_X$ and $SU(3)_C$. 
$f_\text{gf}^{(A)}$, $f_\text{gf}^{(B)}$ and $f_\text{gf}^{(G)}$ are gauge-fixing functions and 
$\xi_{(A)}$, $\xi_{(B)}$ and $\xi_{(G)}$ are gauge parameters. 
$\mathcal{L}_{GH}^{(A)}$, $\mathcal{L}_{GH}^{(B)}$ and $\mathcal{L}_{GH}^{(G)}$ denote ghost Lagrangians, respectively. 
$\Psi_a^{g}$ ($a=1,2,3,4$ and $g=1,2,3$) are the four $SO(5)$-vector (\textbf{5} representation) fermions for each generation 
and $\Psi_{F_i}$ ($i=1,\cdots,N_F$) are the $N_F$ number of $SO(5)$-spinor (\textbf{4} representation) fermions, 
which exist in the bulk space. 
The colour indices are not shown. 
The $SO(5)$ gauge fields $A_M$ are decomposed as 
\begin{equation}
	A_M=\sum^3_{a_L=1}A_M^{a_L}T^{a_L}+\sum^3_{a_R=1}A_M^{a_R}T^{a_R}
	+\sum^4_{\hat{a}=1}A_M^{\hat{a}}T^{\hat{a}} ,
\end{equation}
where $T^{a_L, a_R}  (a_L, a_R = 1, 2, 3)$ and $T^{\hat{a}} (\hat{a} = 1, 2, 3, 4)$ 
are the generators of $SO(4) \simeq SU(2)_L \times SU(2)_R$ and $SO(5)/SO(4)$, respectively. 
For the fermion Lagrangian, 
$\Gamma^M$ denotes 5D gamma matrices which is defined by $\{ \Gamma^M,\Gamma^N \} = 2\eta^{MN}$ ($\eta^{55} = +1$), 
\begin{align}
	&
	\Gamma^{\mu}=\gamma^\mu = 
	\begin{pmatrix} &\sigma^{\mu} \\ \bar\sigma^{\mu}& \end{pmatrix} ~,~~
	\Gamma^5= \gamma^5 = \begin{pmatrix} 1& \\ &-1\end{pmatrix} ~, \nonumber\\
	&
	\sigma^{\mu}=(1, \, \vec{\sigma}) ~,~~
	\bar \sigma ^{\mu}= (-1,\, \vec{\sigma}) ~.
	\label{DiracM}
\end{align}
and $\bar{\Psi}\equiv i\Psi^\dagger \Gamma^0$. 
$e_A{}^M$ is an inverse fielbein and $\Omega_{MBC}$ is the spin connection. 
$\epsilon(y) \equiv \sigma'/k$ is the sign function and $c_a$ are the bulk mass parameters. 
The bulk mass parameters are set as $c_1^g=c_2^g$ and $c_3^g=c_4^g$. 

The $SO(5)$-vector fermions are decomposed to the $SU(2)_L\times SU(2)_R$ bidoublet and singlet. 
The multiplet of the third generation are denoted as
\begin{alignat}{3}
	\Psi_1 &= \left[ \begin{pmatrix} T & t \\ B & b \end{pmatrix}, t' \right]
	&&= \Big[ (Q_1,q),t' \Big]
	&&= \left[ \hat{\Psi}_1, t'\right],
	\nonumber\\
	\Psi_2 &= \left[ \begin{pmatrix} U & X \\ D & Y \end{pmatrix}, b' \right]
	&&= \Big[ (Q_2,Q_3),b' \Big]
	&&= \left[ \hat{\Psi}_2, b'\right],
	\nonumber\\
	\Psi_3 &=\left[ \begin{pmatrix} \nu_\tau & L_{1X} \\ \tau & L_{1Y} \end{pmatrix} , \tau' \right]
	&&= \Big[ (\ell,L_1),\tau' \Big]
	&&= \left[ \hat{\Psi}_3,\tau' \right],
	\nonumber\\
	\Psi_4 &= \left[ \begin{pmatrix} L_{2X} & L_{3X} \\ L_{2Y} & L_{3Y} \end{pmatrix}, \nu'_\tau \right]
	&&= \Big[ (L_2,L_3),\nu'_\tau \Big]
	&&= \left[ \hat{\Psi}_4, \nu'_\tau \right].\label{SO(5)-vector}
\end{alignat}
The first and second generation quarks and leptons are abbreviated. 

The orbifold boundary conditions at $y_0 = 0$ and $y_1 = L$ are given by
\begin{align}
	\begin{pmatrix}A_{\mu}\\A_y\end{pmatrix} (x,y_j-y)&=
	P_\text{ve} \begin{pmatrix}A_{\mu}\\-A_y\end{pmatrix} (x,y_j+y) P_\text{ve}^{-1},  \nonumber\\
	\begin{pmatrix}B_{\mu}\\ B_y\end{pmatrix}(x,y_j-y)&=
	\begin{pmatrix}B_{\mu}\\-B_y\end{pmatrix}(x,y_j+y), \nonumber\\
	\begin{pmatrix}G_{\mu}\\ G_y\end{pmatrix}(x,y_j-y)&=
	\begin{pmatrix}G_{\mu}\\-G_y\end{pmatrix}(x,y_j+y), \nonumber\\
	\Psi_a^g(x,y_j-y)&=P_\text{ve} \Gamma^5 \Psi_a^g(x,y_j+y), \nonumber\\
	\Psi_{F_i} (x, y_j -y)&=(-1)^j P_\text{sp} \Gamma^5\Psi_{F_i} (x, y_j + y), \nonumber\\
	P_\text{ve}=\text{diag} \, ( -1,-1, -1,-1, &+1) , \quad 
	P_\text{sp}=\text{diag} \, ( +1,+1, -1,-1 ) .
	\label{BC}    
\end{align}
By these boundary conditions, 
the $SO(5)\times U(1)_X$ symmetry is broken to $SO(4)\times U(1)_X \simeq SU(2)_L\times SU(2)_R\times U(1)_X$. 
For quarks and leptons, the bidoublets have the left-handed zero modes and the singlets have the right-handed zero modes. 

The brane action is given by 
\begin{align}
S_\text{brane}&= \int d^5 x \sqrt{-G} \, \delta(y) \bigg\{
-(D_\mu \hat\Phi)^\dagger D^\mu \hat\Phi
- \lambda_{\hat \Phi}(\hat \Phi^\dagger \hat \Phi-w^2)^2  \nonumber\\
&+ \sum_{\alpha=1}^3 \big( \hat{\chi}^{q\dagger}_{\alpha R}  \, 
i \bar{\sigma}^\mu D_\mu \hat{\chi}^q_{\alpha R}
+\hat{\chi}^{l\dagger}_{\alpha R}i\bar{\sigma}^\mu D_\mu\hat{\chi}^l_{\alpha R} \big) \nonumber\\
&-i\Bigl
[\kappa_1^q\hat{\chi}_{1R}^{q\dagger}\hat{\Psi}_{1L}\tilde{\hat\Phi}
+\kappa_2^q\hat{\chi}_{2R}^{q\dagger}\hat{\Psi}_{2L}\tilde{\hat\Phi}
+\kappa_3^q\hat{\chi}_{3R}^{q\dagger}\hat{\Psi}_{2L}\hat \Phi
+\tilde{\kappa}^q\hat{\chi}_{2R}^{q\dagger}\hat{\Psi}_{1L} \hat \Phi -(\text{h.c.})\Bigr] \nonumber\\
&-i\Bigl
[\kappa_1^l\hat{\chi}_{1R}^{l\dagger}\hat{\Psi}_{3L}\hat\Phi
+\kappa_2^l\hat{\chi}_{2R}^{l\dagger}\hat{\Psi}_{4L}\tilde{\hat\Phi}
+\kappa_3^l\hat{\chi}_{3R}^{l\dagger}\hat{\Psi}_{4L}\hat \Phi
+\tilde{\kappa}^l\hat{\chi}_{3R}^{l\dagger}\check{\Psi}_{3L}\tilde{\hat\Phi}
-(\text{h.c.}) \Bigr] \bigg\} , 
\label{brane-action}\\
&D_\mu \hat\Phi= \Big( \partial_\mu-ig_A \sum^3_{a_R=1}A^{a_R}_\mu T^{a_R}
-ig_B Q_X B_\mu\Big) \hat \Phi ~, \nonumber\\
&D_\mu\hat \chi_{\alpha R}  =\Big(\partial_\mu -ig_A\sum^3_{a_L=1}A^{a_L}_\mu T^{a_L}
-ig_B Q_{X, \alpha} B_\mu -ig_C Q_{C, \alpha}G_\mu \Big) \hat \chi_{\alpha R} ,\nonumber\\
&\tilde{\hat \Phi}=i\sigma_2 \hat \Phi^* ,\nonumber 
\end{align}
where 
\begin{align}
\hat{\chi}_{1R}^q&= \begin{pmatrix} \hat{T}_R^q \\ \hat{B}_R^q \end{pmatrix}_{7/6},
\quad
\hat{\chi}_{2R}^q = \begin{pmatrix} \hat{U}_R^q \\ \hat{D}_R^q \end{pmatrix}_{1/6},
\quad
\hat{\chi}_{3R}^q = \begin{pmatrix} \hat{X}_R^q \\ \hat{Y}_R^q \end{pmatrix}_{-5/6},
\nonumber\\
\hat{\chi}_{1R}^\ell&= \begin{pmatrix} \hat{L}_{1XR}^\ell \\ \hat{L}_{1YR}^\ell \end{pmatrix}_{-3/2},
\quad
\hat{\chi}_{2R}^\ell = \begin{pmatrix} \hat{L}_{2XR}^\ell \\ \hat{L}_{2YR}^\ell \end{pmatrix}_{1/2},
\quad
\hat{\chi}_{3R}^\ell = \begin{pmatrix} \hat{L}_{3XR}^\ell \\ \hat{L}_{3YR}^\ell \end{pmatrix}_{-1/2},
\label{brane-mass}
\end{align}
are the right-handed fermions localised on $y=0$. 
$\kappa^q_{1,2,3}$, $\kappa^\ell_{1,2,3}$, $\tilde{\kappa}^{q}$ and $\tilde{\kappa}^{\ell}$ are the brane Yukawa couplings. 
The brane scaler breaks the $SU(2)_R \times U(1)_X$ symmetry to the $U(1)_Y$ symmetry spontaneously by acquiring the vacuum expectation value,  
\begin{align}
\langle \Phi \rangle = \begin{pmatrix} 0 \\ w \end{pmatrix}.
\end{align}
The details of the brane interactions are shown in Ref.~\cite{Sakamura:2006rf,Hosotani:2008tx}. 
The formulae of the KK expansion, the mass spectrum of the KK modes and the couplings are shown in Ref~\cite{Funatsu:2016uvi}.

\section{Parameters, couplings and decay widths}
\subsection{Bulk mass parameter dependence}
I would like to emphasize the parameter dependence of fermion mode functions and fermion couplings with the KK gauge boson in this subsection. 
Generally, fermion with the bulk mass parameter $c$ on the warped metric is expanded as~\cite{Grossman:1999ra}
\begin{align}
\Psi_{L,R}(x,y) = \frac{e^{\frac{3}{2}ky}}{\sqrt{L}}\sum_{n=0}^\infty \psi_{L,R}^{(n)}(x)\frac{f_{L,R}^{(n)}(y)}{\sqrt{N^{(n)}}},
\end{align}
where $N^{(n)}$ is the normalisation factor defined as
\begin{align}
N^{(n)}=\int_0^L \frac{dy}{L}\left\{f_{L,R}^{(n)}(y)\right\}^2,
\end{align}
and $f_{L,R}^{(n)}(y)$ is expressed as
\begin{align}
f_{L,R}^{(0)}(y)&=e^{(\frac{1}{2}\mp c)ky}, 
\label{Eq:wave-function1}\\
f_{L,R}^{(n)}(y)&=e^{ky}\left\{ 
a_n J_{c\pm\frac{1}{2}, c\pm\frac{1}{2}} \left(\frac{m_n}{k} e^{ky}\right)
+b_n Y_{c\pm\frac{1}{2}, c\pm\frac{1}{2}} \left(\frac{m_n}{k} e^{ky}\right)\right\}\ \text{for}\ n\neq0,
\label{Eq:wave-function2}
\end{align}
where $a_n$ and $b_n$ are constants determined by boundary conditions.
The upper and lower sign of $\pm$ and $\mp$ correspond to the left- and right-handed fermions, respectively. 
Therefore the left-handed zero-mode is localised towards $y=0$ and $y=L$ for $\frac{1}{2}<c$ and $0<c<\frac{1}{2}$. 
In contrast, the right-handed zero-mode is localised towards $y=L$ for $0<c$. 
These behaviour of the left- and right-handed fermions are reversed for $c<0$. 

From~\eqref{Eq:wave-function1} and \eqref{Eq:wave-function2}, 
it is straightforwardly derived that
the product of the left- and right-handed fermion mode functions with the same KK number and the bulk mass, 
$f_{L}^{(n)}(y)\times f_{R}^{(n)}(y)$ is invariant under changing the sign of the bulk mass parameter $c\to-c$. 
Consequently, the Yukawa couplings and the fermion masses obtained from the Higgs boson vacuum expectation value are independent of the sign of the bulk mass parameters 
in the case that the left- and right- handed quarks have the same bulk mass. 
In this model, the above arguments are applicable.

Considering gauge boson, mode function of zero-mode is independent of $y$-coordinate and 1st KK gauge bosons have peaks near $y=L$~\cite{Pomarol:1999ad}. 
Therefore the right-handed fermions with $0<c$, the left-handed fermions with $c<0$ and the left-handed fermions with $-\frac{1}{2}<c<\frac{1}{2}$ couple to $Z'$ bosons rather largely. 
As shown in the next subsection, the bulk mass parameter of the third generation quarks is $|c_t|<\frac{1}{2}$. 
Therefore the left-handed third generation quarks couplings with $Z'$ bosons are large.

\subsection{Parameters}
\begin{table*}[thbp]
	\centering
	\caption{Input parameters. Masses of $Z$ boson, the Higgs boson, leptons and quarks in the unit of GeV at the $Z$ mass scale. }\label{tbl:inputs}
	\begin{tabular}{cccccc}
		\hline
		$\alpha_\text{EM}^{-1}$ & $\sin^2\theta_W$ & & & & \\
		$127.96$ & $0.23122$ & & & & \\
		\hline
		$m_Z$ & $m_H$ & $m_\nu$ & $m_e$ & $m_\mu$ & $m_\tau$ \\
		$91.188$ & $125$ & $10^{-12}$ & $0.48657 \times 10^{-3}$ & $102.72 \times 10^{-3}$ & $1.7462$ \\
		\hline
		$m_u$ & $m_d$ & $m_s$ & $m_c$ & $m_b$ & $m_t$  \\
		$1.27 \times 10^{-3}$ & $2.90 \times 10^{-3}$ & $0.055$ & $0.619$ & $2.89$ & $171.7$ \\
		\hline
	\end{tabular}
\end{table*}
\begin{table*}[thbp]
	\centering
	\caption{The warp factor, the bulk mass parameters of the fermion and AdS curvature with each values of $\theta_H$ are listed. 
		The resultant $W$-boson mass, Higgs boson mass and KK scale given from the model parameters are also summarized. }
	\begin{tabular}{c|cccccc}
		\hline
		$\theta_H$ & $e^{kL}$ & $\left|c_F\right|$ & $k$ (GeV) & $m_W$ (GeV) & $m_H$ (GeV) & $m_\text{KK}$ (GeV) \\
		\hline
		$0.10$ & $2.90 \times 10^4$ & $0.29617$ & $7.4431 \times 10^7$ & $79.957$ & $125.1$ & $8063$ \\
		$0.09$ & $1.70 \times 10^4$ & $0.27670$ & $4.7190 \times 10^7$ & $79.958$ & $125.1$ & $8721$ \\
		$0.08$ & $1.01 \times 10^4$ & $0.25356$ & $3.0679 \times 10^7$ & $79.951$ & $125.4$ & $9544$ \\
		\hline
		\hline
		$\theta_H$ & $\left|c_e\right|$ & $\left|c_\mu\right|$ & $\left|c_\tau\right|$ & $\left|c_u\right|$ & $\left|c_c\right|$ & $\left|c_t\right|$ \\
		\hline
		$0.10$ & $1.8734$ & $1.3139$ & $1.0060$ & $1.6796$ & $1.1200$ & $0.16116$ \\
		$0.09$ & $1.9504$ & $1.3599$ & $1.0348$ & $1.7459$ & $1.1552$ & $0.11646$ \\
		$0.08$ & $2.0342$ & $1.4100$ & $1.0663$ & $1.8180$ & $1.1936$ & $0.0089140$ \\
		\hline
	\end{tabular}\label{tbl:outputs}
\end{table*}
{The free parameters of this model are the warp factor $e^{kL}$ and $N_F$ which is $\Psi_F$'s degrees of the freedom, 
so once the $e^{kL}$ and $N_F$ are set, $\theta_H$ is determined. 
The physics of the quarks and leptons are almost independent of $N_F$ and determined by $\theta_H$~\cite{Funatsu:2013ni,Funatsu:2014fda}. }
The input parameters and the model parameters to realise the input parameters at the tree level are listed in Table~\ref{tbl:inputs} and 
Table~\ref{tbl:outputs} respectively, 
where $c_e$, $c_\mu$, $c_\tau$, $c_u$, $c_c$, $c_t$ and $c_F$ are the bulk mass parameters of leptons and quarks for the each generations and $\Psi_F$. 
{As explained in previous subsection, 
the sign of the bulk mass parameter $c$ doesn't affect the fermion mass in this model. }
Therefore only the absolute values of $c$ is determined from the input values. 
In the following, the bulk mass parameters of leptons and quarks are abbreviated as 
$c_l\equiv (c_e, c_\mu, c_\tau)$ $c_q\equiv (c_u, c_c, c_t)$. 
The resultant $W$-boson mass at the tree-level calculated by the boundary condition is $m_W^\text{tree}=80.0$~GeV. 
To realise the input parameters, the parameter region of $\theta_H$ is found to be $0.08\le \theta_H \le 0.10$. 
The lower limit of $\theta_H$ becomes slightly smaller for $N_F>4$. 
In $N_F=8$ case, lower limit of $\theta_H=0.078$.

\subsection{Couplings and decay widths}
\renewcommand{\arraystretch}{0.8}
\begin{table*}[tbp]
	\footnotesize 
	\centering
	\caption{
		Couplings of neutral vector bosons ($Z'$ bosons) to fermions in unit of $g_w = e/\sin\theta_W$ for $\theta_H = 0.10$, $c_l$, $c_q>0$. }
	\label{tbl:coupling_10p}
	\begin{tabular}{c|cc|cc|cc|cc}
		$f$ & $g^L_{Zf}$ & $g^R_{Zf}$ & $g^L_{Z^{(1)}f}$ & $g^R_{Z^{(1)}f}$ & $g^L_{Z_R^{(1)}}$ & $g^R_{Z_R^{(1)}f}$ & $g^L_{\gamma^{(1)}f}$ & $g^R_{\gamma^{(1)}f}$ \\
		\hline
		$\nu_e$    & $+0.57037$ & $0$ & $-0.20943$ & $0$ & $0$ & $0$ & $0$ & $0$ \\
		$\nu_\mu$  & $+0.57037$ & $0$ & $-0.20943$ & $0$ & $0$ & $0$ & $0$ & $0$ \\
		$\nu_\tau$ & $+0.57037$ & $0$ & $-0.20928$ & $0$ & $0$ & $0$ & $0$ & $0$ \\
		\hline
		$e$    & $-0.30661$ & $+0.26384$ & $+0.11258$ & $+1.04332$ & $0$ & $-1.4357$ & $+0.17720$ & $-1.8962$ \\
		$\mu$  & $-0.30661$ & $+0.26384$ & $+0.11258$ & $+0.97948$ & $0$ & $-1.3582$ & $+0.17720$ & $-1.7801$ \\
		$\tau$ & $-0.30661$ & $+0.26383$ & $+0.11250$ & $+0.92684$ & $0$ & $-1.2940$ & $+0.17708$ & $-1.6844$ \\
		\hline 
		$u$ & $+0.39453$ & $-0.17589$ & $-0.14486$ & $-0.68311$ & $0$ & $+0.94208$ & $-0.11813$ & $+1.2415$ \\
		$c$ & $+0.39453$ & $-0.17589$ & $-0.14485$ & $-0.63219$ & $0$ & $+0.88013$ & $-0.11812$ & $+1.1489$ \\
		$t$ & $+0.39353$ & $-0.17694$ & $+0.57109$ & $-0.42117$ & $+1.1369$ & $+0.62142$ & $+0.46722$ & $+0.76730$ \\
		\hline
		$d$ & $-0.48245$ & $+0.087946$ & $+0.17715$ & $+0.34156$ & $0$ & $-0.47104$ & $+0.059066$ & $-0.62077$ \\
		$s$ & $-0.48245$ & $+0.087945$ & $+0.17713$ & $+0.31609$ & $0$ & $-0.44006$ & $+0.059060$ & $-0.57445$ \\
		$b$ & $-0.48252$ & $+0.087939$ & $-0.70659$ & $+0.21112$ & $+1.1347$ & $-0.31045$ & $-0.23377$ & $-0.38353$ \\
	\end{tabular}
%
\vspace{1em}
%
	\footnotesize
	\centering
	\caption{
		$Z'$ couplings of fermions for $\theta_H = 0.09$, $c_l$, $c_q>0$. The same unit as in Table~\ref{tbl:coupling_10p}.}
	\label{tbl:coupling_09p}
	\begin{tabular}{c|cc|cc|cc|cc}
		$f$ & $g^L_{Zf}$ & $g^R_{Zf}$ & $g^L_{Z^{(1)}f}$ & $g^R_{Z^{(1)}f}$ & $g^L_{Z_R^{(1)}}$ & $g^R_{Z_R^{(1)}f}$ & $g^L_{\gamma^{(1)}f}$ & $g^R_{\gamma^{(1)}f}$ 
		\\
		\hline
		$\nu_e$      & $+0.57035$ & 0 & $-0.21569$ & 0 & 0 & 0 & 0 & 0 \\
		$\nu_{\mu}$  & $+0.57035$ & 0 & $-0.21569$ & 0 & 0 & 0 & 0 & 0 \\
		$\nu_{\tau}$ & $+0.57035$ & 0 & $-0.21553$ & 0 & 0 & 0 & 0 & 0 \\
		\hline
		$e$    & $-0.30660$ & $+0.26382$ & $+0.11595$ & $+1.02101$ & 0 & $-1.4062$ & $+0.18238$ & $-1.8568$ \\
		$\mu$  & $-0.30660$ & $+0.26382$ & $+0.11595$ & $+0.95843$ & 0 & $-1.3307$ & $+0.18238$ & $-1.7430$ \\
		$\tau$ & $-0.30660$ & $+0.26382$ & $+0.11586$ & $+0.90600$ & 0 & $-1.2671$ & $+0.18225$ & $-1.6476$ \\
		\hline
		$u$ & $+0.39452$ & $-0.17588$ & $-0.14919$ & $-0.66857$ & $0$ & $+0.92287$ & $-0.12159$ & $+1.2159$ \\
		$c$ & $+0.39452$ & $-0.17588$ & $-0.14918$ & $-0.61829$ & $0$ & $+0.86208$ & $-0.12157$ & $+1.1244$ \\
		$t$ & $+0.39363$ & $-0.17681$ & $+0.61316$ & $-0.39018$ & $+1.2038$ & $+0.58325$& $+0.50090$ & $+0.71132$ \\
		\hline
		$d$ & $-0.48244$ & $+0.087940$ & $+0.18244$ & $+0.33428$ & $0$ & $-0.46144$ & $+0.060793$ & $-0.60793$ \\
		$s$ & $-0.48244$ & $+0.087939$ & $+0.18242$ & $+0.30915$ & $0$ & $-0.43104$ & $+0.060787$ & $-0.56219$ \\
		$b$ & $-0.48250$ & $+0.087933$ & $-0.75660$ & $+0.19561$ & $+1.2016$ & $-0.29141$ & $-0.25054$ & $-0.35559$ \\
	\end{tabular}
%
\vspace{1em}
%
	\footnotesize
	\centering
	\caption{
		$Z'$ couplings of fermions for $\theta_H = 0.08$, $c_l$, $c_q>0$.
		The same unit as in Table~\ref{tbl:coupling_10p}.}
	\label{tbl:coupling_08p}
	\begin{tabular}{c|cc|cc|cc|cc}
		$f$ & $g^L_{Zf}$ & $g^R_{Zf}$ & $g^L_{Z^{(1)}f}$ & $g^R_{Z^{(1)}f}$ & $g^L_{Z_R^{(1)}}$ & $g^R_{Z_R^{(1)}f}$ & $g^L_{\gamma^{(1)}f}$ & $g^R_{\gamma^{(1)}f}$ 
		\\
		\hline
		$\nu_e$      & $+0.57034$ & 0 & $-0.22233$ & 0 & 0 & 0 & 0 & 0 \\
		$\nu_{\mu}$  & $+0.57034$ & 0 & $-0.22233$ & 0 & 0 & 0 & 0 & 0 \\
		$\nu_{\tau}$ & $+0.57034$ & 0 & $-0.22216$ & 0 & 0 & 0 & 0 & 0 \\
		\hline
		$e$    & $-0.30659$ & $+0.26380$ & $+0.11952$ & $+0.99861$ & 0 & $-1.3762$ & $+0.18789$ & $-1.8171$ \\
		$\mu$  & $-0.30659$ & $+0.26380$ & $+0.11952$ & $+0.93739$ & 0 & $-1.3029$ & $+0.18789$ & $-1.7057$ \\
		$\tau$ & $-0.30659$ & $+0.26380$ & $+0.11943$ & $+0.88524$ & 0 & $-1.2401$ & $+0.18775$ & $-1.6107$ \\
		\hline
		$u$ & $+0.39451$ & $-0.17587$ & $-0.15379$ & $-0.65398$ & $0$ & $+0.90342$ & $-0.12526$ & $+1.1900$ \\
		$c$ & $+0.39451$ & $-0.17587$ & $-0.15377$ & $-0.60444$ & $0$ & $+0.84393$ & $-0.12524$ & $+1.0998$ \\
		$t$ & $+0.39365$ & $-0.17675$ & $+0.71984$ & $-0.33029$ & $+1.3690$ & $+0.50941$ & $+0.50941$ & $+0.58717$ \\
		\hline
		$d$ & $-0.48242$ & $+0.087934$ & $+0.18806$ & $+0.32699$ & $0$ & $-0.45171$ & $+0.062629$ & $-0.59500$ \\
		$s$ & $-0.48242$ & $+0.087934$ & $+0.18804$ & $+0.30222$ & $0$ & $-0.43197$ & $+0.062622$ & $-0.54990$ \\
		$b$ & $-0.48248$ & $+0.087927$ & $-0.88581$ & $+0.16571$ & $+1.3666$ & $-0.25452$ & $-0.29359$ & $-0.30139$ \\
	\end{tabular}
\end{table*}
\begin{table*}[tbp]
	\footnotesize
	\centering
	\caption{
		$Z'$ couplings of fermions for $\theta_H = 0.10$, $c_l$, $c_q<0$.
		The same unit as in Table~\ref{tbl:coupling_10p}.}
	\label{tbl:coupling_10n}
	\begin{tabular}{c|cc|cc|cc|cc}
		$f$ & $g^L_{Zf}$ & $g^R_{Zf}$ & $g^L_{Z^{(1)}f}$ & $g^R_{Z^{(1)}f}$ & $g^L_{Z_R^{(1)}}$ & $g^R_{Z_R^{(1)}f}$ & $g^L_{\gamma^{(1)}f}$ & $g^R_{\gamma^{(1)}f}$ 
		\\
		\hline
		$\nu_e$    & $+0.57054$ & $0$ & $+2.2561$ & $0$ & $-3.1047$ & $0$ & $0$ & $0$ \\
		$\nu_\mu$  & $+0.57053$ & $0$ & $+2.1181$ & $0$ & $-2.9371$ & $0$ & $0$ & $0$ \\
		$\nu_\tau$ & $+0.57052$ & $0$ & $+2.0042$ & $0$ & $-2.7982$ & $0$ & $0$ & $0$ \\
		\hline
		$e$    & $-0.30384$ & $+0.26662$ & $-1.2015$ & $-0.09790$ & $-3.1130$ & $0$ & $-1.8962$ & $+0.17720$ \\
		$\mu$  & $-0.30384$ & $+0.26662$ & $-1.1280$ & $-0.09790$ & $-2.9450$ & $0$ & $-1.7801$ & $+0.17720$ \\
		$\tau$ & $-0.30383$ & $+0.26662$ & $-1.0674$ & $-0.09783$ & $-2.8057$ & $0$ & $-1.6844$ & $+0.17708$ \\
		\hline 
		$u$ & $+0.39419$ & $-0.17630$ & $+1.5309$ & $+0.06473$ & $-1.3595$ & $0$ & $+1.2415$ & $-0.11813$ \\
		$c$ & $+0.39180$ & $-0.17868$ & $+1.4082$ & $+0.06560$ & $+2.3878$ & $0$ & $+1.1489$ & $-0.11812$ \\
		$t$ & $+0.39281$ & $-0.17766$ & $+0.9400$ & $-0.25520$ & $+1.7082$ & $+0.41431$ & $+0.7673$ & $+0.46722$ \\
		\hline
		$d$ & $-0.48019$ & $+0.09032$ & $-1.8649$ & $-0.03316$ & $-1.3614$ & $0$ & $-0.62077$ & $+0.05907$ \\
		$s$ & $-0.48256$ & $+0.08794$ & $-1.7344$ & $-0.03229$ & $+2.3805$ & $0$ & $-0.57445$ & $+0.05906$ \\
		$b$ & $-0.48255$ & $+0.08793$ & $-1.1585$ & $+0.12877$ & $+1.7027$ & $-0.20689$ & $-0.38353$ & $-0.23377$ \\
	\end{tabular}
%
\vspace{1em}
%
	\footnotesize
	\centering
	\caption{
		$Z'$ couplings of fermions for $\theta_H = 0.09$, $c_l$, $c_q<0$
		The same unit as in Table~\ref{tbl:coupling_10p}.}
	\label{tbl:coupling_09n}
	\begin{tabular}{c|cc|cc|cc|cc}
		$f$ & $g^L_{Zf}$ &  $g^R_{Zf}$ & $g^L_{Z^{(1)}f}$ & $g^R_{Z^{(1)}f}$ & $g^L_{Z_R^{(1)}}$ & $g^R_{Z_R^{(1)}f}$ & $g^L_{\gamma^{(1)}f}$ & $g^R_{\gamma^{(1)}f}$ 
		\\
		\hline
		$\nu_e$      & $+0.57050$ & 0 & $+2.2079$ & 0 & $-3.0408$ & 0 & 0 & 0 \\
		$\nu_{\mu}$  & $+0.57050$ & 0 & $+2.0725$ & 0 & $-2.8775$ & 0 & 0 & 0 \\
		$\nu_{\tau}$ & $+0.57048$ & 0 & $+1.9592$ & 0 & $-2.7400$ & 0 & 0 & 0 \\
		\hline
		$e$    & $-0.30436$ & $+0.26607$ & $-1.1779$ & $-0.10062$ & $-3.0474$ & $0$ & $-1.8568$ & $+0.18238$ \\
		$\mu$  & $-0.30436$ & $+0.26607$ & $-1.1057$ & $-0.10062$ & $-2.8838$ & $0$ & $-1.7430$ & $+0.18238$ \\
		$\tau$ & $-0.30436$ & $+0.26607$ & $-1.0452$ & $-0.10055$ & $-2.7460$ & $0$ & $-1.6476$ & $+0.18225$ \\
		\hline
		$u$ & $+0.39424$ & $-0.17621$ & $+1.4986$ & $+0.06664$ & $-1.3314$ & $0$ & $+1.2159$ & $-0.12159$ \\
		$c$ & $+0.39231$ & $-0.17813$ & $+1.3792$ & $+0.06736$ & $+2.3375$ & $0$ & $+1.1244$ & $-0.12157$ \\
		$t$ & $+0.39322$ & $-0.17723$ & $+0.8717$ & $-0.27396$ & $+1.6023$ & $+0.43863$ & $+0.7113$ & $+0.50090$ \\
		\hline
		$d$ & $-0.48061$ & $+0.08986$ & $-1.8270$ & $-0.03398$ & $-1.3329$ & $0$ & $-0.60793$ & $+0.06079$ \\
		$s$ & $-0.48253$ & $+0.08794$ & $-1.6963$ & $-0.03325$ & $+2.3317$ & $0$ & $-0.56219$ & $+0.06079$ \\
		$b$ & $-0.48252$ & $+0.08793$ & $-1.0734$ & $+0.13788$ & $+1.5982$ & $-0.21909$ & $-0.35559$ & $-0.25054$ \\
	\end{tabular}
%
\vspace{1em}
%
	\footnotesize
	\centering
	\caption{
		$Z'$ couplings of fermions for $\theta_H = 0.08$, $c_l$, $c_q<0$
		The same unit as in Table~\ref{tbl:coupling_10p}.}
	\label{tbl:coupling_08n}
	\begin{tabular}{c|cc|cc|cc|cc}
		$f$ & $g^L_{Zf}$ & $g^R_{Zf}$ & $g^L_{Z^{(1)}f}$ & $g^R_{Z^{(1)}f}$ & $g^L_{Z_R^{(1)}}$ & $g^R_{Z_R^{(1)}f}$ & $g^L_{\gamma^{(1)}f}$ & $g^R_{\gamma^{(1)}f}$ 
		\\
		\hline
		$\nu_e$      & $+0.57046$ & 0 & $+2.1594$ & 0 & $-2.9760$ & 0 & 0 & 0 \\
		$\nu_{\mu}$  & $+0.57045$ & 0 & $+2.0271$ & 0 & $-2.8174$ & 0 & 0 & 0 \\
		$\nu_{\tau}$ & $+0.57045$ & 0 & $+1.9143$ & 0 & $-2.6816$ & 0 & 0 & 0 \\
		\hline
		$e$    & $-0.30483$ & $+0.26557$ & $-1.1539$ & $-0.10353$ & $-2.9811$ & $0$ & $-1.8171$ & $+0.18789$ \\
		$\mu$  & $-0.30482$ & $+0.26557$ & $-1.0832$ & $-0.10353$ & $-2.8223$ & $0$ & $-1.7057$ & $+0.18789$ \\
		$\tau$ & $-0.30482$ & $+0.26557$ & $-1.0229$ & $-0.10345$ & $-2.6862$ & $0$ & $-1.6107$ & $+0.18775$ \\
		\hline
		$u$ & $+0.39429$ & $-0.17612$ & $+1.4626$ & $+0.06866$ & $-1.3031$ & $0$ & $+1.1900$ & $-0.12526$ \\
		$c$ & $+0.39277$ & $-0.17764$ & $+1.3499$ & $+0.06924$ & $+2.2871$ & $0$ & $+1.0998$ & $-0.12524$ \\
		$t$ & $+0.39363$ & $-0.17678$ & $+0.7390$ & $-0.32167$ & $+1.3984$ & $+0.49872$ & $+0.6028$ & $+0.5872$ \\
		\hline
		$d$ & $-0.48099$ & $+0.08945$ & $-1.7886$ & $-0.03487$ & $-1.3042$ & $0$ & $-0.59500$ & $+0.06269$ \\
		$s$ & $-0.48250$ & $+0.08793$ & $-1.6583$ & $-0.03427$ & $+2.2826$ & $0$ & $-0.54990$ & $+0.06262$ \\
		$b$ & $-0.48248$ & $+0.08793$ & $-0.9093$ & $+0.16143$ & $+1.3959$ & $-0.24918$ & $-0.30139$ & $-0.29359$ \\
	\end{tabular}
\end{table*}
\renewcommand{\arraystretch}{1}
The fermion couplings to $Z'$s are listed in the Table~\ref{tbl:coupling_10p}, \ref{tbl:coupling_09p}, \ref{tbl:coupling_08p}, \ref{tbl:coupling_10n}, \ref{tbl:coupling_09n} and \ref{tbl:coupling_08n}. 
The 1st KK photon couplings to the left- and right-handed fermions for $c>0$ is equal to the left- and right-handed fermions for $c<0$ within 5 digits, respectively. 
The $Z_R^{(1)}$ coupling with ($u_L$, $d_L$) for $c<0$ is negative, 
although the couplings with ($c_L$, $s_L$) and ($t_L$, $b_L$) for $c<0$ are positive. 
This behaviour comes from the mass ratio of up-type and down-type quarks. 
In this model, left-handed top-quark is the mixing state of the ($t_L$, $B_L$, $U_L$) in~\eqref{SO(5)-vector}
and their mixing ratio is
\begin{align}
( t_L : B_L : U_L) = \left(\frac{1+c_H}{2} : \frac{1-c_H}{2}  : -\frac{\tilde{\kappa}}{\kappa_2}\right).
\end{align}
where $\tilde{\kappa}/\kappa_2 \simeq m_b/m_t$ in good accuracy. 
Their $SU(2)_L \times SU(2)_R$ isospins are 
\begin{align}
t_L: \left(+\frac{1}{2}, -\frac{1}{2}\right) ,\quad
B_L: \left(-\frac{1}{2}, +\frac{1}{2}\right) ,\quad
U_L: \left(+\frac{1}{2}, +\frac{1}{2}\right) 
\end{align}
Therefore up-type quark coupling with the $U(1)_X$ gauge boson is proportional to 
\begin{align}
\left(\frac{1+c_H}{2}\right)^2 + \left(\frac{1-c_H}{2}\right)^2  + \left(\frac{m_b}{m_t}\right)^2,
\end{align}
up-type quark coupling with the $SU(2)_L$ neutral gauge boson is proportional to 
\begin{align}
\left(\frac{1+c_H}{2}\right)^2 - \left(\frac{1-c_H}{2}\right)^2  + \left(\frac{m_b}{m_t}\right)^2,
\end{align}
and up-type quark coupling with the $SU(2)_R$ neutral gauge boson is proportional to 
\begin{align}-
\left(\frac{1+c_H}{2}\right)^2 + \left(\frac{1-c_H}{2}\right)^2  + \left(\frac{m_b}{m_t}\right)^2.
\end{align}
For small $\theta_H$, the sign of the last one depends on the ratio of the up- and down-type quarks. 
Thus the $Z_R^{(1)}$ couplings with the up-quark have different sign with that of the charm- and top-quarks 
The behaviour of the $Z_R^{(1)}$ couplings with the down-type quarks are derived by the same reason.

\begin{table}[htp]
	\centering
	\caption{
		Masses of $Z^{(1)}$, $Z_R^{(1)}$ and $\gamma^{(1)}$ and total decay width of $\gamma^{(1)}$ in the unit of GeV. 
		$\Gamma_{\gamma^{(1)}}$ is independent of the sign of the bulk fermion parameters. 
		$\Gamma_{Z^{(1)}/Z_R^{(1)}}(\pm,\pm)$ represent that left and right sign is sign of $c_l$ and $c_q$.}
	\label{tbl:masses}
	\begin{tabular}{c|cccc}
		\hline
		$\theta_H$ & $m_{Z^{(1)}}$ & $m_{Z_R^{(1)}}$ & $m_{\gamma^{(1)}}$ & $\Gamma_{\gamma^{(1)}}$ \\ 
		\hline
		$0.10$ & $6585$ & $6172$ & $6588$ & $905$ \\
		$0.09$ & $7149$ & $6676$ & $7152$ & $940$ \\
		$0.08$ & $7855$ & $7305$ & $7858$ & $986$ \\
		\hline 
		\hline
		$\theta_H$ 
		& $\Gamma_{Z^{(1)}}(+,+)$ 
		& $\Gamma_{Z^{(1)}}(+,-)$ 
		& $\Gamma_{Z^{(1)}}(-,+)$ 
		& $\Gamma_{Z^{(1)}}(-,-)$ 
		\\ 
		\hline
		$0.10$ & $429$ & $1632$ & $959$  & $2162$ \\
		$0.09$ & $463$ & $1674$ & $1014$ & $2225$ \\
		$0.08$ & $534$ & $1705$ & $1112$ & $2283$ \\
		\hline 
		\hline
		$\theta_H$ 
		& $\Gamma_{Z_R^{(1)}}(+,+)$ 
		& $\Gamma_{Z_R^{(1)}}(+,-)$ 
		& $\Gamma_{Z_R^{(1)}}(-,+)$ 
		& $\Gamma_{Z_R^{(1)}}(-,-)$ 
		\\ 
		\hline
		$0.10$ & $784$  & $2437$ & $2398$ & $4051$ \\
		$0.09$ & $856$  & $2480$ & $2529$ & $4153$ \\
		$0.08$ & $1005$ & $2485$ & $2758$ & $4238$ \\
		\hline 
	\end{tabular}
\end{table}
The $Z'$ masses obtained by the boundary conditions and the decay widths calculated from the couplings shown in
Table~\ref{tbl:coupling_10p}, \ref{tbl:coupling_09p}, \ref{tbl:coupling_08p}, \ref{tbl:coupling_10n}, \ref{tbl:coupling_09n} and \ref{tbl:coupling_08n}
are summarized in Table~\ref{tbl:masses}. 
The 1st KK photon decay width is independent of the sign of the bulk fermion parameters.

\section{Cross section and forward-backward asymmetry}
The parameters are constrained by the experimental results of the forward-backward asymmetry at the $Z$-pole. 
For $c_l>0$, the deviations of the $Z$-boson couplings are $O(0.01) \%$. 
In contrast, for $c_l<0$, their deviations are $O(0.1) \%$. 
Thus the forward-backward asymmetry of $e^+e^-\to \mu^+\mu^-$ process at the $Z$-pole in the GHU model deviate nearly 10 \% from the observed value for $c_l<0$. 
Consequently, the value $\sin^2\theta_W=0.23122$ is not valid and 
the value of $\theta_W$ which consistently explain the experimental results at the $Z$-pole must be searched. 
In this paper, $c_l<0$ case is not considered further.

The longitudinal polarisation $P_{e^\pm}$ ($-1 \le P_{e^\pm} \le 1$) is introduced, where  
the electron and positron is purely right-handed when $P_{e^\pm}=1$. 
The cross section of $e^-e^+ \to Z' \to f\bar{f}$ at the centre-of-mass frame is given by
\begin{align}
\frac{d\sigma}{d\cos\theta} &=
\frac{1}{4} \bigg[ (1-P_{e^-})(1+P_{e^+}) \frac{d\sigma_{LR}}{d\cos\theta} 
+ (1+P_{e^-})(1-P_{e^+}) \frac{d\sigma_{RL}}{d\cos\theta} \bigg],
\label{eq:sigma}
\end{align}
where $\sigma_{LR}$ ($\sigma_{RL}$) is $e_L^-e_R^+ (e_R^- e_L^+) \to f\bar{f}$ cross section. 
The formula \eqref{eq:sigma} is rewritten by using $P_\text{eff} = (P_{e^-} - P_{e^+})/ (1 - P_{e^-} P_{e^+})$ as 
$\sigma(P_\text{eff},0) = \sigma(P_{e^-},P_{e^+})/(1 - P_{e^-}P_{e^+})$, 
then the ratio of $\sigma$ is parametrised by one polarisation parameter $P_\text{eff}$. 
Typical values of polarisation parameters are $(P_{e^-},P_{e^+}) = (\pm 0.8,\mp 0.3)$ 
($P_\text{eff} = \pm 0.887$). 

Considering the $e^+e^-\to \mu^+\mu^-$ process, 
the difference of the cross sections 
between $c_q>0$ case ($\sigma^{c_q>0}$) and $c_q<0$ case ($\sigma^{c_q<0}$) arises from only the $Z'$ decay widths. 
Consequently the deviation of $\sigma^{c_q<0}(\mu^+\mu^-)$ from $\sigma^{c_q>0}(\mu^+\mu^-)$ is small. 
As shown in ~\cite{Funatsu:2017nfm}, 
at $\sqrt{s} = 250$ GeV with 250 $\text{fb}^{-1}$ unpolarised beam, 
$4.66 \times 10^5$ events are expected in the SM. 
Therefore the statistical uncertainty is 0.15~\%. 
The difference of the cross sections of the two cases over the SM value, 
$(\sigma^{c_q>0}-\sigma^{c_q<0})/\sigma^{\text{SM}}(\mu^+\mu^-)$ is less than 0.11\% at $\sqrt{s}=250$ GeV with unpolarised beam. 
For the forward-backward asymmetry, 
$(A_\text{FB}^{c_q>0}-A_\text{FB}^{c_q<0})/A_\text{FB}^{\text{SM}}(\mu^+\mu^-)$ is less than 0.04 \%
at $\sqrt{s}=250$ GeV with unpolarised beam. 
Thus the two cases are difficult to distinguish at the $e^+e^-\to \mu^+\mu^-$ process. 
The detailed analysis of the $e^+e^-\to \mu^+\mu^-$ process in the GHU model is shown in~\cite{Funatsu:2017nfm}. 

\begin{figure}[htb]
	\centering
	\includegraphics[width=0.95\linewidth]{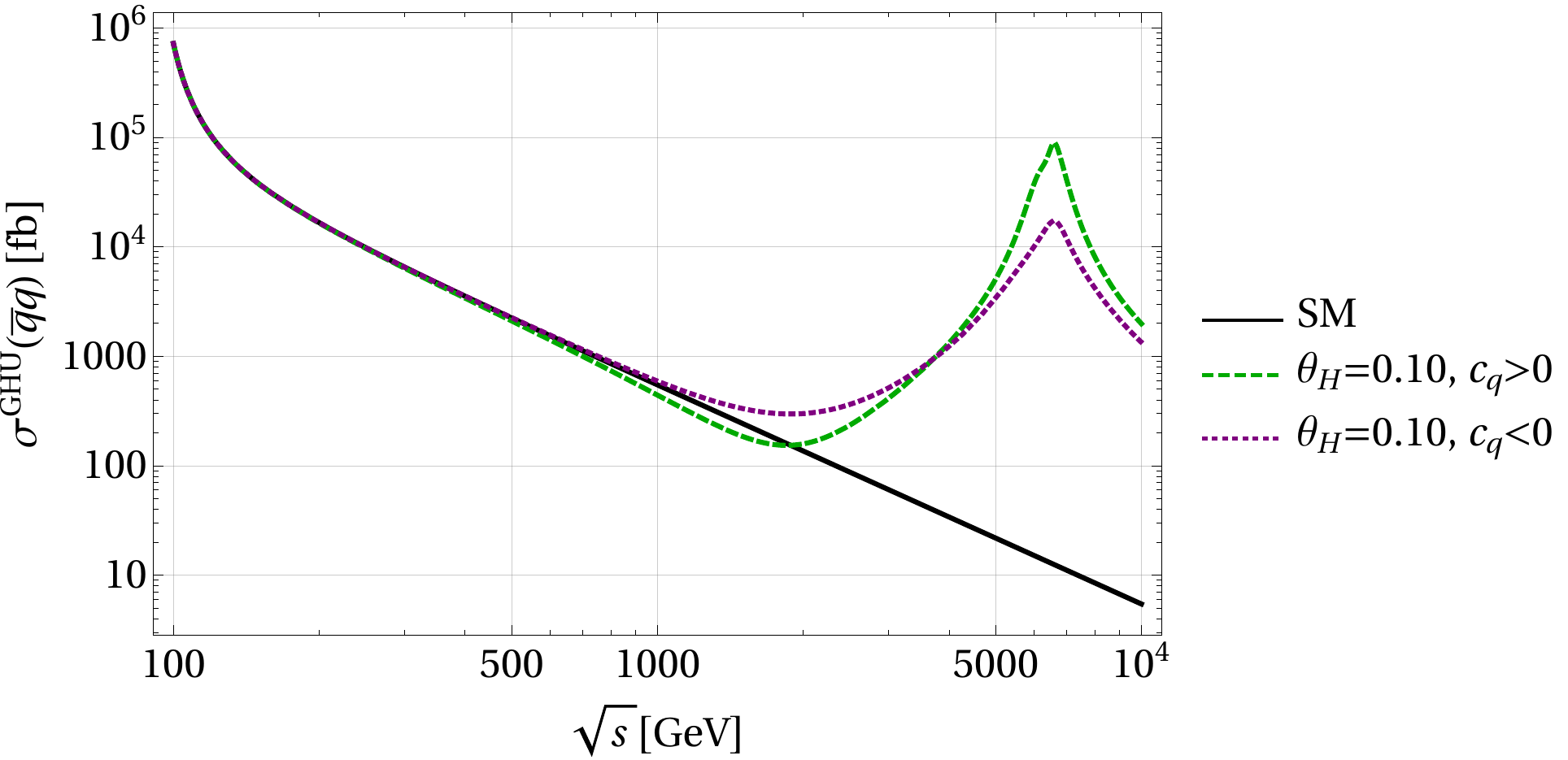}
	\caption{
		Cross sections of the $e^+e^-\to \bar{q}q~(q\neq b, t)$ process with unpolarised beams in the SM and GHU model for $\theta_H=0.10$. 
		Black solid line represents the cross sections in the SM. 
		Green dashed and purple dotted lines correspond to that in GHU model with $c_q>0$ and $c_q<0$, respectively. }
	\label{fig:sigmaqq-10TeV}
\end{figure}
\begin{figure}[H]\centering
	\begin{minipage}{0.45\hsize}
		\centering
		\includegraphics[width=0.95\linewidth]{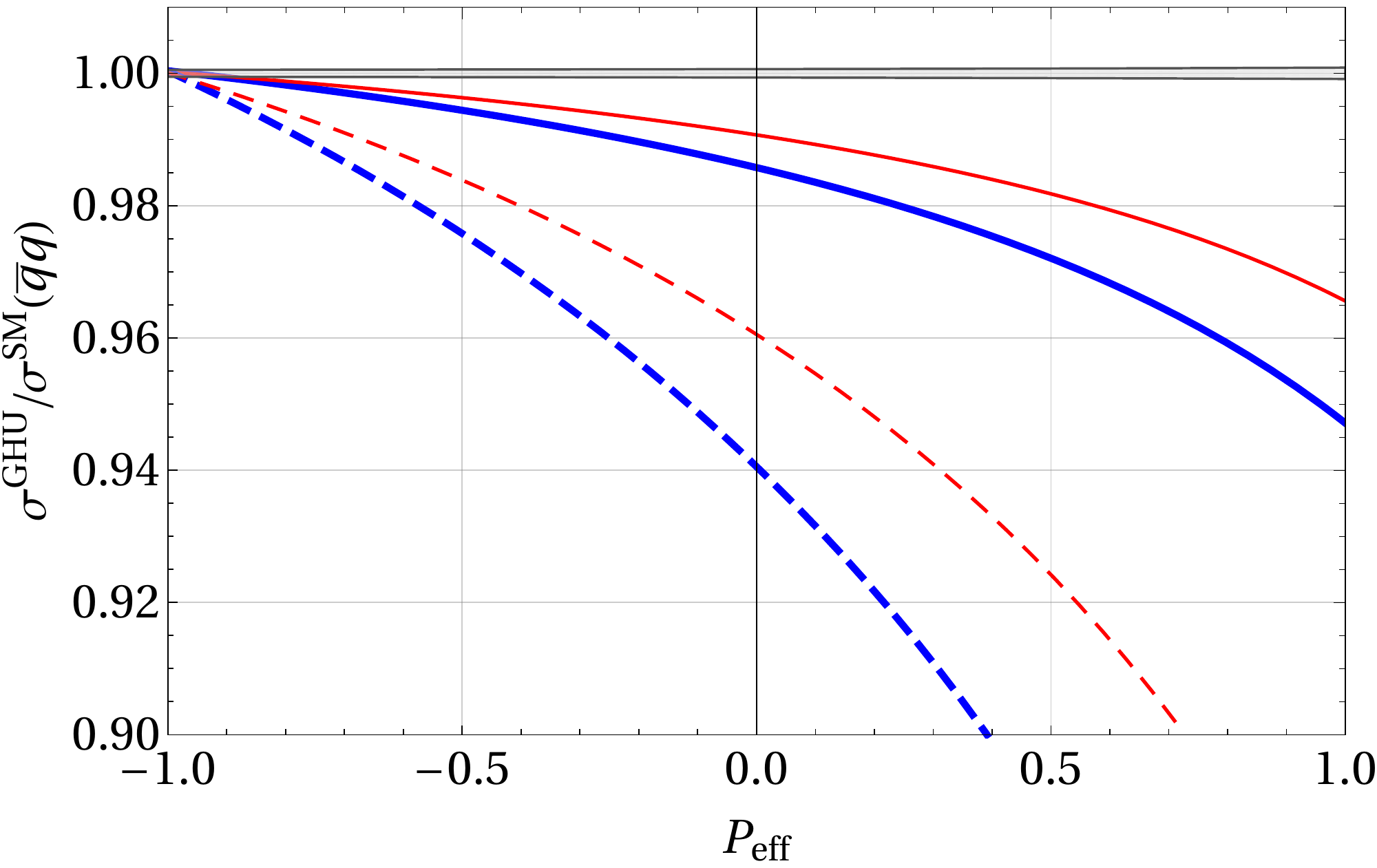}
		\subcaption{$\sigma^\text{GHU}/\sigma^\text{SM}(\bar{q}q)$ for $c_q>0$ case. }
	\end{minipage}
\hspace{0.04\hsize}
	\begin{minipage}{0.45\hsize}
		\centering
		\includegraphics[width=0.95\linewidth]{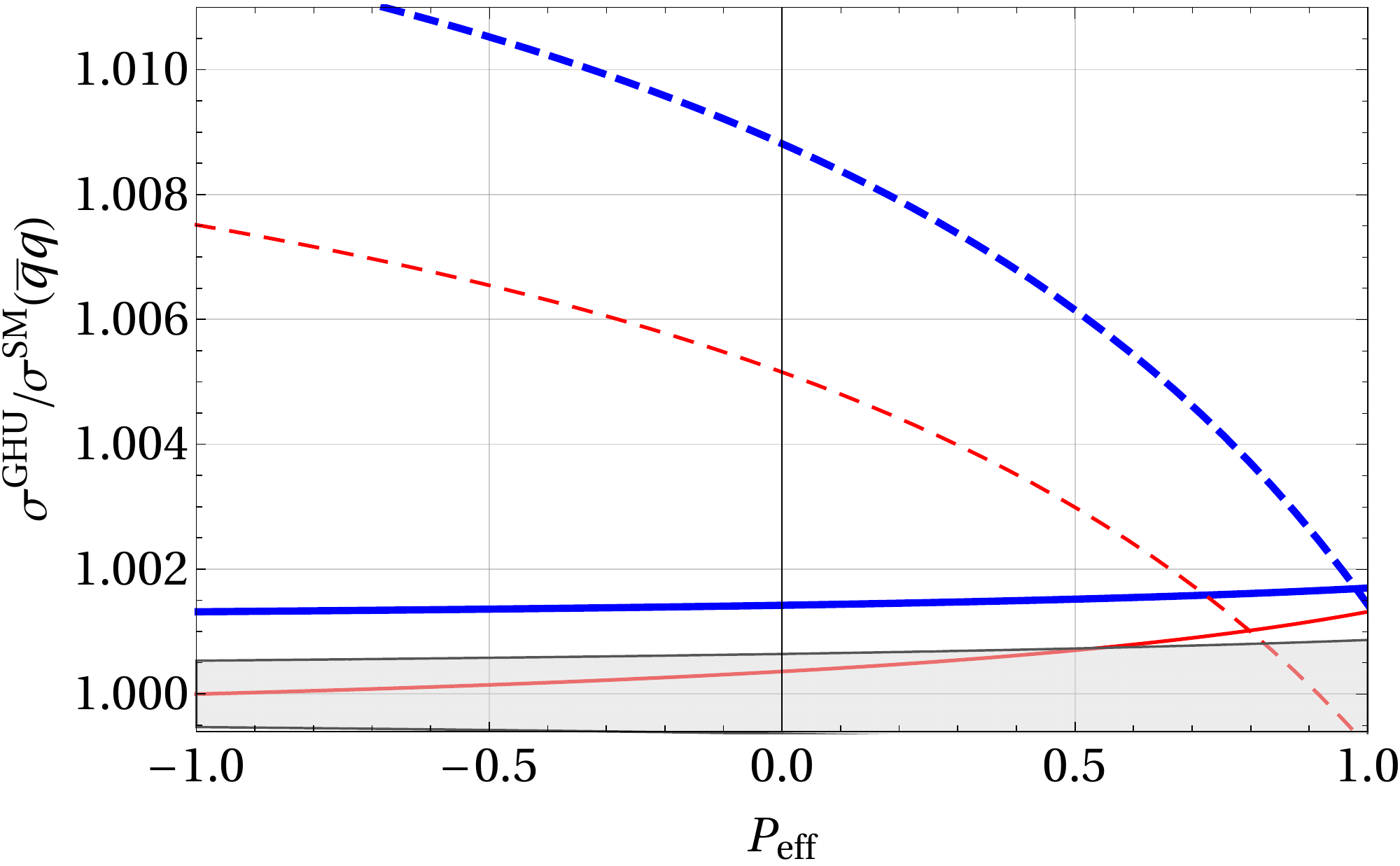}
		\subcaption{$\sigma^\text{GHU}/\sigma^\text{SM}(\bar{q}q)$ for $c_q<0$ case. }
	\end{minipage}
	\caption{
		Ratio of the cross sections in the GHU model to that in the SM with polarisation beams for the $e^+e^-\to \bar{q}q ~(q\neq b, t)$ process. 
		The left figure shows the $c_q>0$ case and the right figure shows the $c_q<0$ case. 
		Solid and dotted lines are for $\sqrt{s} = 250$ GeV and $500$ GeV, respectively.
		Blue-thick and red-thin lines correspond to $\theta_H = 0.10$ and $0.08$, respectively.
		The gray band indicates the  statistical uncertainty at $\sqrt{s}=250$ GeV with $250\text{ fb}^{-1}$ data.}
	\label{fig:sigmaqq}
\end{figure}
For the $e^+e^-\to \bar{q}q ~(q\neq b, t)$ process, 
the cross section in the SM is $\sigma^\text{SM}(\bar{q}q)$= 9.75~pb and 7.26~pb at $\sqrt{s}=250$ GeV with unpolarised and polarised ($P_{e^-}=+0.8$ and $P_{e^+}=-0.3$) beams, respectively. 
For $\sqrt{s}=O(100)$ GeV region, the cross section in the GHU model with $c_q>0$ are smaller than that in the SM. 
In contrast, that in the GHU model with $c_q<0$ are larger than that in the SM as shown in figure~\ref{fig:sigmaqq-10TeV}. 
In figure~\ref{fig:sigmaqq}, the ratio of cross sections in the GHU model to that in the SM with polarised beams are plotted. 
At $\sqrt{s}=250$ GeV with $P_\text{eff}=+0.887$ and 250 fb$^{-1}$ luminosity, 
the event number of the SM is $1.814\times 10^6$ and the statistical uncertainty $\sigma$ over the event number is $0.074\%$. 
$\sigma(\bar{q}q)$ in the GHU model deviates from the that in the SM larger than 3 \% for $c_q>0$ 
at $\sqrt{s}=250$ GeV with $P_\text{eff}=+0.887$ and 250 fb$^{-1}$ luminosity. 
The deviations at $\sqrt{s} = 250$ GeV are summarised in Table~\ref{tbl:sigma250}. 
\begin{table}[thbp]
	\centering
	\caption{
		Deviations of the cross sections at $\sqrt{s}=250$ GeV with $P_{e^-}=+0.8$ and $P_{e^+}=-0.3$ and 250 fb$^{-1}$ luminosity
		for $e^+e^-\to \bar{q}q ~(q\neq b, t)$ and $e^+e^-\to \bar{b}b$ processes. 
		The statistical uncertainties calculated by the SM prediction are 0.074 \% for $\bar{q}q$ and 0.20 \% for $\bar{b}b$.}
	\label{tbl:sigma250}
	\begin{tabular}{c|cc}
		\hline
		$\theta_H$ & 
		$\sigma^{c_q>0}/\sigma^\text{SM}(\bar{q}q)-1$ & 
		$\sigma^{c_q<0}/\sigma^\text{SM}(\bar{q}q)-1$ \\ 
		\hline
		$0.10$ & $-4.56 \%$~($-61.5\sigma$) & $+0.16 \%$~($+2.21\sigma$) \\
		$0.09$ & $-3.70 \%$~($-49.9\sigma$) & $+0.14 \%$~($+1.90\sigma$) \\
		$0.08$ & $-2.98 \%$~($-40.0\sigma$) & $+0.11 \%$~($+1.53\sigma$) \\
		\hline 
		\hline
		$\theta_H$ & 
		$\sigma^{c_q>0}/\sigma^\text{SM}(\bar{b}b)-1$ & 
		$\sigma^{c_q<0}/\sigma^\text{SM}(\bar{b}b)-1$ \\ 
		\hline
		$0.10$ & $-4.18 \%$~($-21.1\sigma$)  & $-3.96 \%$~($-19.9\sigma$)  \\
		$0.09$ & $-3.41 \%$~($-17.2\sigma$)  & $-3.29 \%$~($-16.6\sigma$)  \\
		$0.08$ & $-2.84 \%$~($-14.3\sigma$)  & $-2.65 \%$~($-13.3\sigma$)  \\
		\hline 
	\end{tabular}
\end{table}
\begin{figure}[ht]\centering
	\begin{minipage}{0.45\hsize}
		\centering
		\includegraphics[width=0.95\linewidth]{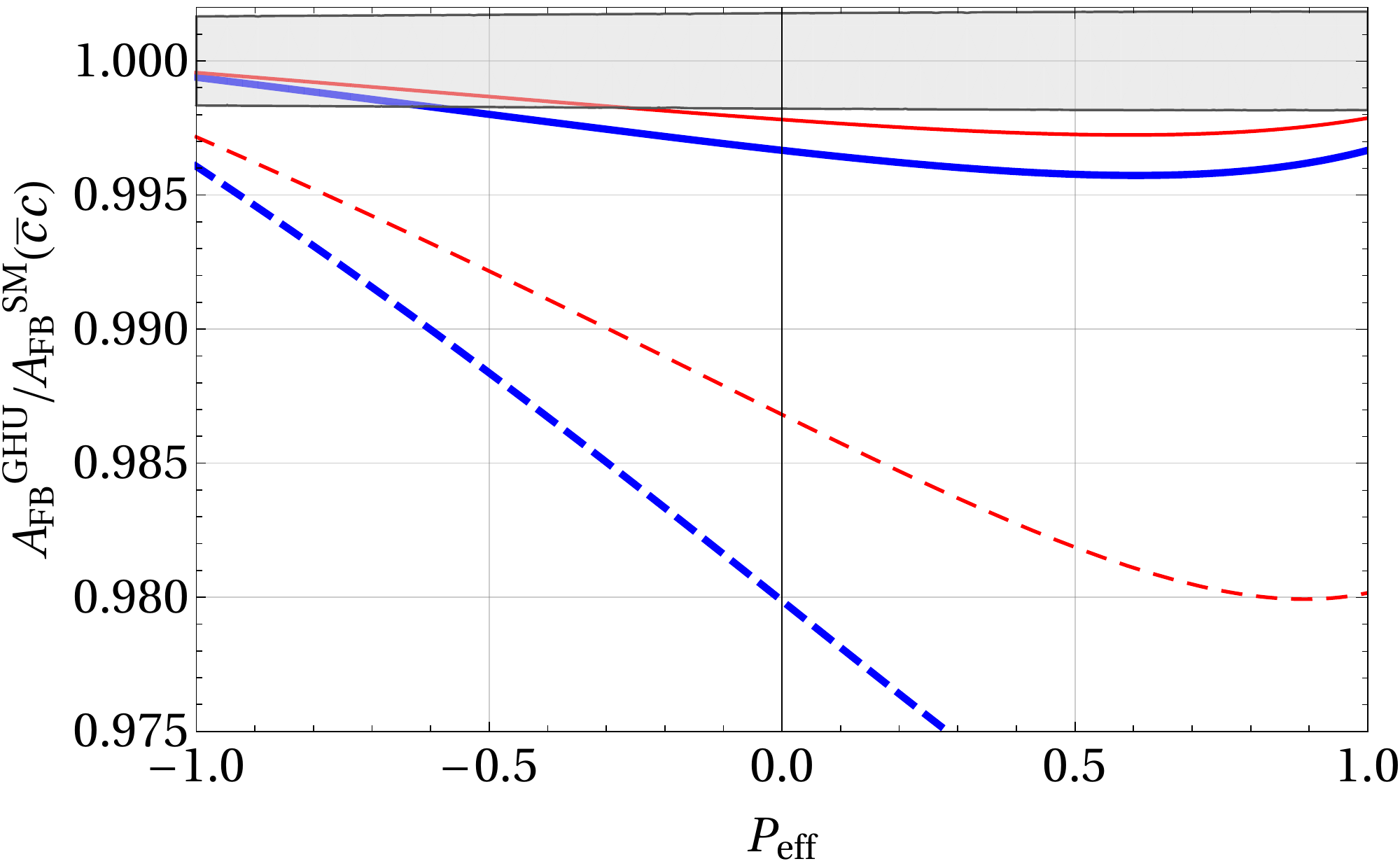}
		\subcaption{$A_\text{FB}^\text{GHU}/A_\text{FB}^\text{SM}(\bar{c}c)$ for $c_q>0$ case. }
	\end{minipage}
\hspace{0.04\hsize}
	\begin{minipage}{0.45\hsize}
		\centering
		\includegraphics[width=0.95\linewidth]{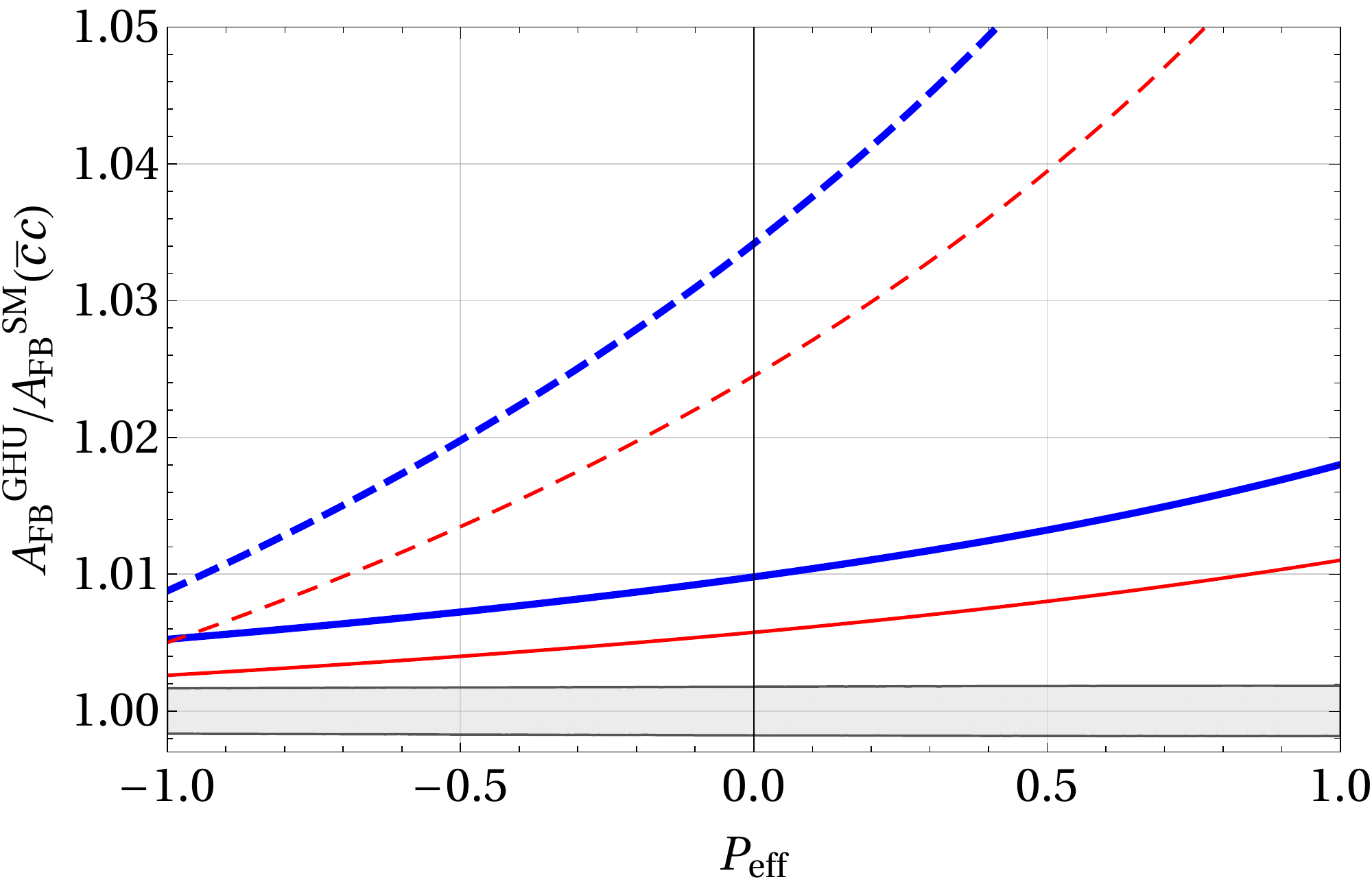}
		\subcaption{$A_\text{FB}^\text{GHU}/A_\text{FB}^\text{SM}(\bar{c}c)$ for $c_q<0$ case. }
	\end{minipage}
	\caption{		
		Ratio of the forward-backward asymmetry in the GHU model to that in the SM with polarisation beams for the $e^+e^-\to \bar{c}c$ process. 
		The left figure shows the $c_q>0$ case and the right figure shows the $c_q<0$ case. 
		Solid and dotted lines are for $\sqrt{s} = 250$ GeV and $500$ GeV, respectively.
		Blue-thick and red-thin lines correspond to $\theta_H = 0.10$ and $0.08$, respectively.
		The gray band indicates the statistical uncertainty at $\sqrt{s}=250$ GeV with $250\text{ fb}^{-1}$ data.}
	\label{fig:AFBcc}
\end{figure}
For the $\bar{c}c$ final state, the forward-backward asymmetry is also measured. 
The $A_\text{FB}^\text{GHU}(\bar{c}c)$ decrease from the $A_\text{FB}^\text{SM}(\bar{c}c)$ for $c_q>0$. 
In contrast, $A_\text{FB}^\text{GHU}(\bar{c}c)$ increase from the $A_\text{FB}^\text{SM}(\bar{c}c)$ for $c_q<0$. 
At $\sqrt{s}=250$ GeV with $P_\text{eff}=+0.887$ polarised beam, 
$A_\text{FB}^\text{SM}(\bar{c}c)=0.700$ and with 250 $\text{fb}^{-1}$ luminosity, the statistical uncertainty is 0.16\%. 
The deviations of the $A_\text{FB}^\text{GHU}(\bar{c}c)$ from $A_\text{FB}^\text{SM}(\bar{c}c)$ are $-0.38\%$, $-0.31\%$ and $-0.25\%$ 
for $c_q>0$ and $\theta_H=0.10$, $0.09$ and $0.08$ at $\sqrt{s}=250$ GeV with $P_\text{eff}=+0.887$. 
For $c_q<0$ and $\theta_H=0.10$, $0.09$ and $0.08$ at $\sqrt{s}=250$ GeV with $P_\text{eff}=+0.887$, 
the deviations are $+1.68\%$, $+1.25\%$ and $+1.03\%$. 
The polarisation dependence of the ratio $A_\text{FB}^\text{GHU}/A_\text{FB}^\text{SM}(\bar{c}c)$ are plotted in figure~\ref{fig:AFBcc}
and the deviations at $\sqrt{s} = 250$~GeV, 500~GeV and 1~TeV are summarised in Table~\ref{tbl:AFB250}, Table~\ref{tbl:AFB500} and Table~\ref{tbl:AFB1000}, respectively. 
\begin{table}[H]\vspace{-0.5em}
	\centering
	\caption{
		Deviation of $A_\text{FB}(\bar{c}c)$ and $A_\text{FB}(\bar{b}b)$ at $\sqrt{s}=250$ GeV with $P_\text{eff}=+0.887$ and 250 fb$^{-1}$ luminosity. 
		The statistical uncertainties $\sigma$ calculated by the event number of the SM prediction are 0.16~\% and 0.70~\%, respectively. }
	\label{tbl:AFB250}
	\begin{tabular}{c|cc}
		\hline
		$\theta_H$ & 
		$A_\text{FB}^{c_q>0}/A_\text{FB}^\text{SM}(\bar{c}c)-1$ & 
		$A_\text{FB}^{c_q<0}/A_\text{FB}^\text{SM}(\bar{c}c)-1$ \\ 
		\hline
		$0.10$ & $-0.38 \%~(-2.32\sigma)$ & $+1.68 \%~(+10.1\sigma)$ \\
		$0.09$ & $-0.31 \%~(-1.88\sigma)$ & $+1.25 \%~(+7.58\sigma)$ \\
		$0.08$ & $-0.25 \%~(-1.49\sigma)$ & $+1.03 \%~(+6.21\sigma)$ \\
		\hline 
		\hline
		$\theta_H$ & 
		$A_\text{FB}^{c_q>0}/A_\text{FB}^\text{SM}(\bar{b}b)-1$ & 
		$A_\text{FB}^{c_q<0}/A_\text{FB}^\text{SM}(\bar{b}b)-1$ \\ 
		\hline
		$0.10$ & $+4.24 \%~(+6.04\sigma)$ & $+7.33 \%~(+10.4\sigma)$ \\
		$0.09$ & $+4.07 \%~(+5.81\sigma)$ & $+5.69 \%~(+8.10\sigma)$ \\
		$0.08$ & $+4.15 \%~(+5.92\sigma)$ & $+3.83 \%~(+5.45\sigma)$ \\
		\hline 
	\end{tabular}
\end{table}

\begin{figure}[htb]\centering
	\begin{minipage}{0.45\hsize}
		\centering
		\includegraphics[width=0.95\linewidth]{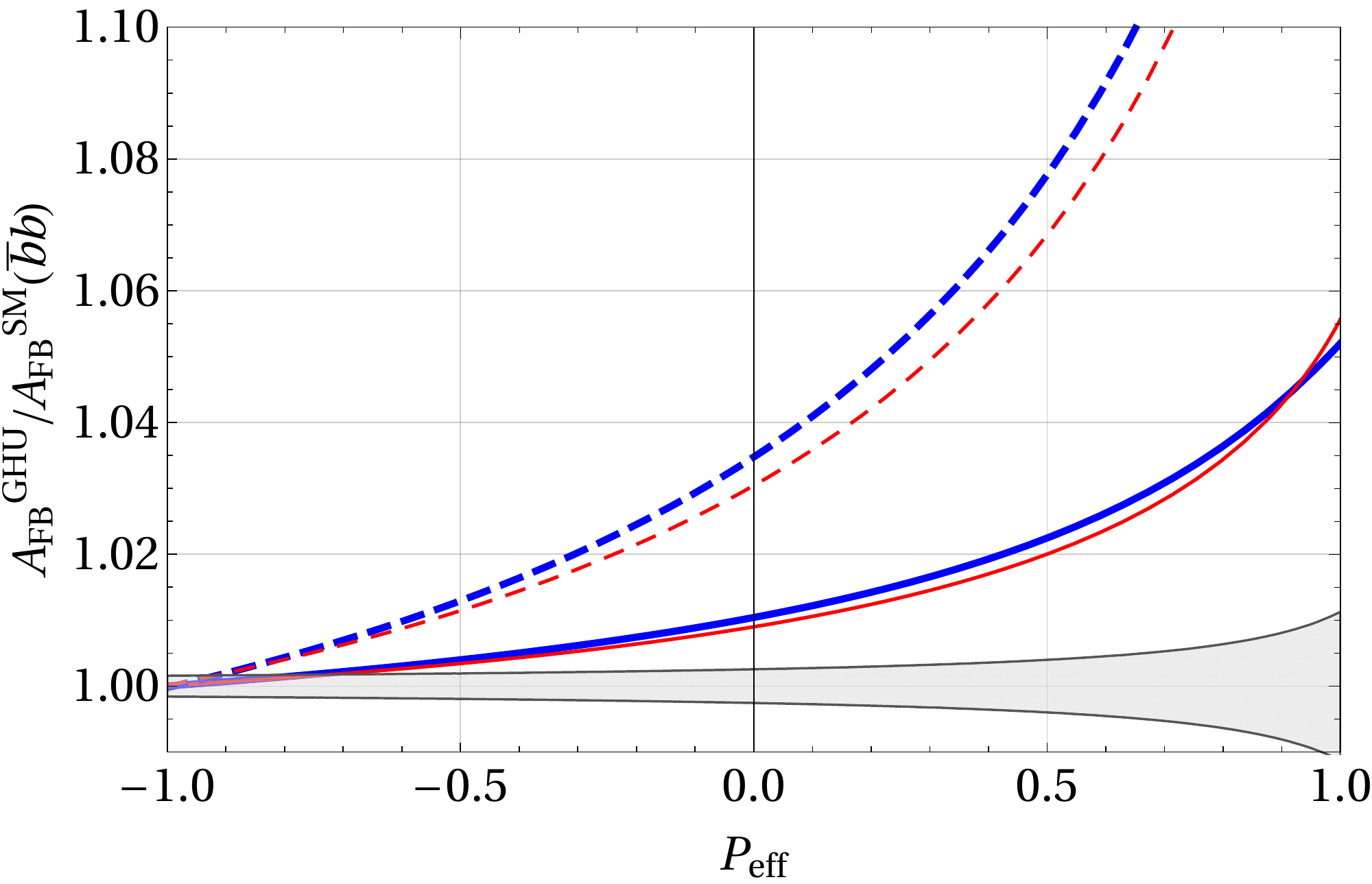}
		\subcaption{$A_\text{FB}^\text{GHU}/A_\text{FB}^\text{SM}(\bar{b}b)$ for $c_q>0$ case. }
	\end{minipage}
\hspace{0.04\hsize}
	\begin{minipage}{0.45\hsize}
		\centering
		\includegraphics[width=0.95\linewidth]{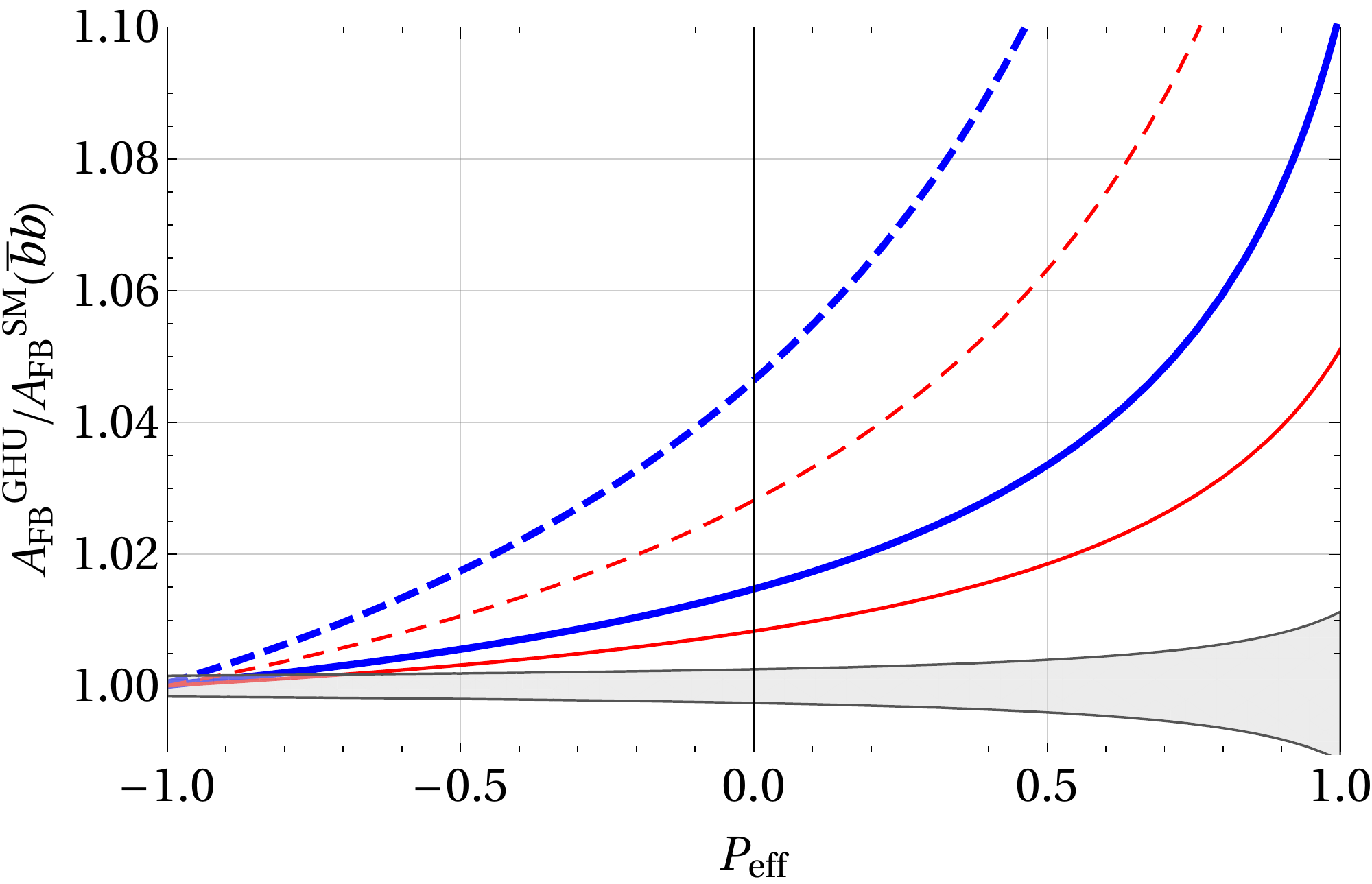}
		\subcaption{$A_\text{FB}^\text{GHU}/A_\text{FB}^\text{SM}(\bar{b}b)$ for $c_q<0$ case. }
	\end{minipage}
	\caption{
		Ratio of the forward-backward asymmetry in the GHU model to that in the SM with polarisation beams for the $e^+e^-\to \bar{b}b$ process. 
		The left figure shows the $c_q>0$ case and the right figure shows the $c_q<0$ case. 
		Solid and dotted lines are for $\sqrt{s} = 250$ GeV and $500$ GeV, respectively.
		Blue-thick and red-thin lines correspond to $\theta_H = 0.10$ and $0.08$, respectively.
		The gray band indicates the statistical uncertainty at $\sqrt{s}=250$ GeV with $250\text{ fb}^{-1}$ data.
		}
	\label{fig:AFBbb}
\end{figure} 
For the $e^+e^-\to \bar{b}b$ process, 
the cross section in the SM is $\sigma^\text{SM}(\bar{b}b)$= 1.77~pb and 1.02~pb at $\sqrt{s}=250$~GeV with unpolarised and polarised ($P_{e^-}=+0.8$ and $P_{e^-}=-0.3$) beams, respectively. 
The statistical uncertainty at $\sqrt{s}=250$~GeV and 250~fb$^{-1}$ luminosity with $P_{e^-}=+0.8$ and $P_{e^+}=-0.3$ beam are 0.20 \%.
For this process, the cross sections in the GHU model with $c_q>0$ and $c_q<0$ cases both decrease from that in the SM. 
The deviations at $\sqrt{s} = 250$~GeV are summarised in Table~\ref{tbl:sigma250}. 
The deference between the $c_q>0$ and $c_q<0$ cases more obviously appear at the forward-backward asymmetry. 
In the SM, $A_\text{FB}^\text{SM}(\bar{b}b)=0.618$ and $0.366$ at $\sqrt{s}=250$~GeV with unpolarised and $P_\text{eff}=+0.887$ beams, respectively. 
In the GHU model the $A_\text{FB}(\bar{b}b)$ increase from the SM value, 
$A_\text{FB}^{c_q>0}(\bar{b}b)$ increase $4.24 \%$, $4.07 \%$, $4.15 \%$ and 
$A_\text{FB}^{c_q<0}(\bar{b}b)$ increase $7.33 \%$, $5.69 \%$, $3.83 \%$ 
at $\sqrt{s}=250$~GeV with $P_\text{eff}=+0.887$ for $\theta_H=0.10$, $0.09$ and $0.08$. 
In figure~\ref{fig:AFBbb}, the ratio of the $A_\text{FB}(\bar{b}b)$ in the GHU model to that in the SM with polarised beams are plotted. 
The $c_q<0$ case predicts larger deviation than the $c_q>0$ case for $\theta_H=0.10$ and $0.09$. 
At $\sqrt{s}=250$~GeV with $P_\text{eff}=+0.887$ and 250~fb$^{-1}$ luminosity, 
the statistical uncertainty of the $A_\text{FB}(\bar{b}b)$ in the SM is 0.70\%. 
$A_\text{FB}(\bar{b}b)$ in the GHU model deviates from the that in the SM larger than $5.4\sigma$. 
The deviations at $\sqrt{s} = 250$~GeV, 500~GeV and 1~TeV are summarised in Table~\ref{tbl:AFB250}, Table~\ref{tbl:AFB500} and Table~\ref{tbl:AFB1000}, respectively.  

The forward-backward asymmetry of the $e^+e^-\to \bar{t}t$ process is measured at $\sqrt{s}=500$~GeV. 
The polarisation dependence of $A_\text{FB}^\text{GHU}/A_\text{FB}^\text{SM}(\bar{t}t)$ is qualitatively similar to that of $A_\text{FB}^\text{GHU}/A_\text{FB}^\text{SM}(\bar{b}b)$, which also increase from that in the SM. 
$A_\text{FB}^{c_q>0}(\bar{t}t)/A_\text{FB}^\text{SM}(\bar{t}t)-1=5.36 \%$, $5.03 \%$, $5.14 \%$ and 
$A_\text{FB}^{c_q<0}(\bar{t}t)/A_\text{FB}^\text{SM}(\bar{t}t)-1=9.25 \%$, $7.23 \%$, $4.91 \%$ at $\sqrt{s}=500$ GeV with $P_\text{eff}=+0.887$ for $\theta_H=0.10$, $0.09$ and $0.08$. 
At $\sqrt{s}=500$~GeV with $P_\text{eff}=+0.887$ and 500~fb$^{-1}$ luminosity, 
$\sigma^\text{SM}(\bar{t}t)=479$ $\text{fb}^{-1}$, $A_\text{FB}^\text{SM}(\bar{t}t)=0.463$
and the uncertainty of the $A_\text{FB}^\text{SM}(\bar{t}t)$ is 0.538 \%.  
The deviations of the forward-backward asymmetries of the $e^+e^-\to \bar{c}c,\ \bar{b}b,\ \bar{t}t$ processes at $\sqrt{s}=500$~GeV and 1~TeV are summarised in Table~\ref{tbl:AFB500} and Table~\ref{tbl:AFB1000}. 

\begin{table}[htp]
	\centering
	\caption{
		Deviations of $A_\text{FB}(\bar{c}c)$, $A_\text{FB}(\bar{b}b)$ and $A_\text{FB}(\bar{t}t)$ at $\sqrt{s}=500$~GeV with $P_\text{eff}=+0.887$ and 500~fb$^{-1}$ luminosity. 
		The statistical uncertainties calculated by the event number of the SM prediction are 0.26 \%, 0.79 \% and 0.53 \%, respectively. }
	\label{tbl:AFB500}
	\begin{tabular}{c|cc}
		\hline
		$\theta_H$ & 
		$A_\text{FB}^{c_q>0}/A_\text{FB}^\text{SM}(\bar{c}c)-1$ & 
		$A_\text{FB}^{c_q<0}/A_\text{FB}^\text{SM}(\bar{c}c)-1$ \\ 
		\hline
		$0.10$ & $-3.25 \%~(-12.3\sigma)$ & $+7.49 \%~(+28.3\sigma)$ \\
		$0.09$ & $-2.58 \%~(-9.77\sigma)$ & $+6.50 \%~(+24.6\sigma)$ \\
		$0.08$ & $-2.01 \%~(-7.59\sigma)$ & $+5.53 \%~(+20.9\sigma)$ \\
		\hline 
		\hline
		$\theta_H$ & 
		$A_\text{FB}^{c_q>0}/A_\text{FB}^\text{SM}(\bar{b}b)-1$ & 
		$A_\text{FB}^{c_q<0}/A_\text{FB}^\text{SM}(\bar{b}b)-1$ \\ 
		\hline
		$0.10$ & $+15.4 \%~(+19.4\sigma)$ & $+23.1 \%~(+29.1\sigma)$ \\
		$0.09$ & $+14.3 \%~(+18.1\sigma)$ & $+18.4 \%~(+23.3\sigma)$ \\
		$0.08$ & $+14.1 \%~(+17.8\sigma)$ & $+12.9 \%~(+16.3\sigma)$ \\
		\hline 
		\hline
		$\theta_H$ & 
		$A_\text{FB}^{c_q>0}/A_\text{FB}^\text{SM}(\bar{t}t)-1$ & 
		$A_\text{FB}^{c_q<0}/A_\text{FB}^\text{SM}(\bar{t}t)-1$ \\ 
		\hline
		$0.10$ & $+5.36 \%~(+9.96\sigma)$ & $+9.25 \%~(+17.2\sigma)$ \\
		$0.09$ & $+5.03 \%~(+9.35\sigma)$ & $+7.23 \%~(+13.4\sigma)$ \\
		$0.08$ & $+5.14 \%~(+9.55\sigma)$ & $+4.91 \%~(+9.13\sigma)$ \\
		\hline
	\end{tabular}
\end{table}

\begin{table}[htp]
	\centering
	\caption{
		Deviation of $A_\text{FB}(\bar{c}c)$, $A_\text{FB}(\bar{b}b)$ and $A_\text{FB}(\bar{t}t)$ at $\sqrt{s}=1$~TeV with $P_\text{eff}=+0.887$ and 1000~fb$^{-1}$ luminosity. 
		The statistical uncertainties calculated by the event number of the SM prediction are 0.39 \%, 1.07 \% and 0.45 \%, respectively. }
	\label{tbl:AFB1000}
	\begin{tabular}{c|cc}
		\hline
		$\theta_H$ & 
		$A_\text{FB}^{c_q>0}/A_\text{FB}^\text{SM}(\bar{c}c)-1$ & 
		$A_\text{FB}^{c_q<0}/A_\text{FB}^\text{SM}(\bar{c}c)-1$ \\ 
		\hline
		$0.10$ & $-25.9 \%~(-67.3\sigma)$ & $-2.46 \%~(-6.37\sigma)$ \\
		$0.09$ & $-18.3 \%~(-47.6\sigma)$ & $+4.61 \%~(+12.0\sigma)$ \\
		$0.08$ & $-12.8 \%~(-33.3\sigma)$ & $+9.02 \%~(+23.4\sigma)$ \\
		\hline 
		\hline
		$\theta_H$ & 
		$A_\text{FB}^{c_q>0}/A_\text{FB}^\text{SM}(\bar{b}b)-1$ & 
		$A_\text{FB}^{c_q<0}/A_\text{FB}^\text{SM}(\bar{b}b)-1$ \\ 
		\hline
		$0.10$ & $+46.1 \%~(+43.1\sigma)$ & $+21.1 \%~(+19.7\sigma)$ \\
		$0.09$ & $+45.7 \%~(+42.8\sigma)$ & $+36.8 \%~(+34.4\sigma)$ \\
		$0.08$ & $+44.9 \%~(+42.0\sigma)$ & $+38.8 \%~(+36.3\sigma)$ \\
		\hline 
		\hline
		$\theta_H$ & 
		$A_\text{FB}^{c_q>0}/A_\text{FB}^\text{SM}(\bar{t}t)-1$ & 
		$A_\text{FB}^{c_q<0}/A_\text{FB}^\text{SM}(\bar{t}t)-1$ \\ 
		\hline
		$0.10$ & $+11.3 \%~(+24.3\sigma)$ & $+9.36 \%~(+20.6\sigma)$ \\
		$0.09$ & $+11.6 \%~(+25.7\sigma)$ & $+12.3 \%~(+27.1\sigma)$ \\
		$0.08$ & $+12.2 \%~(+26.9\sigma)$ & $+11.3 \%~(+25.0\sigma)$ \\
		\hline
	\end{tabular}
\end{table}

\section{Summary}
In the above calculations, 
the quark bulk mass parameters $(c_u, c_c, c_t)$ are assumed to be all positive or all negative. 
It is also allowed to be only one of them is positive or negative. 
In the case, the $Z'$ decay widths change from the values shown in Table~\ref{tbl:masses}, 
therefore the cross sections and the forward-backward asymmetries slightly change from the results in this paper. 
Neglecting the difference arising from the $Z'$ decay widths, 
the sign of the $c_c$ is determined by measuring the $A_\text{FB}(\bar{c}c)$ 
and the sign of the $c_u$ is determined by the $A_\text{FB}(\bar{c}c)$ and $\sigma(\bar{q}q)$ at $\sqrt{s}=250$~GeV. 
It is difficult to determine the sign of the $c_t$ by measuring $A_\text{FB}(\bar{b}b)$ and $A_\text{FB}(\bar{t}t)$. 
At the ILC~250~GeV, 
the $c_q<0$ case predicts $4 \sigma$ larger deviation of the $A_\text{FB}(\bar{b}b)$ than the $c_q<0$ case for $\theta_H=0.10$, 
and at the ILC~500~GeV $5 \sigma$ larger deviation for $\theta_H=0.09$. 
For $\theta_H=0.08$, to clearly determine the sign of $c_t$ by observing the $A_\text{FB}(\bar{t}t)$, 
higher energy and luminosity such as the ILC~1~TeV are necessary. 

In this paper, the forward-backward asymmetry of the $e^+e^-\to \bar{q}q$ processes in the GHU model are studied 
for two cases where all of the quark bulk mass parameters are positive or negative. 
The GHU model predicts large deviations at the $\sqrt{s}=250$~GeV with polarised beams. 
Therefore the GHU model is testable at the ILC~250~GeV. 
The signs of the bulk mass parameters are distinguished at the ILC~500~GeV or ILC~1~TeV. 
For the case where the lepton bulk mass parameters are negative, detail is going to be analysed in near future.

\subsection*{Acknowledgements}
I thank Yutaka Hosotani, Hisaki Hatanaka, Yuta Orikasa and Naoki Yamatsu for many advices and important discussions. 
This work is supported by
the National Natural Science Foundation of China (Grant Nos.~11775092, 11675061, 11521064 and 11435003), 
and the International Postdoctoral Exchange Fellowship Program (IPEFP).

\end{document}